\documentclass[a4paper,11pt]{article}
\pdfoutput=1 

\usepackage{jheppub} 

\usepackage{graphicx}
\usepackage{subfig}
\usepackage[T1]{fontenc} 

\usepackage{slashed}			
 
\title{\boldmath \boldmath Light Mediators in Anomaly Free $U(1)_X$ Models II - Constraints on Dark Gauge Bosons }

\author[a,1]{F.C. Correia,}
\author[b,c,2]{Svjetlana Fajfer}

\affiliation[a]{Institut f\" ur Physik, Technische Universit\" at Dortmund, D-44221 Dortmund, Germany}
\affiliation[b]{Department of Physics, University of Ljubljana, Jadranska 19, 1000 Ljubljana, Slovenia}
\affiliation[c]{J. Stefan Institute, Jamova 39, P. O. Box 3000, 1001 Ljubljana, Slovenia}

\emailAdd{fagner.correia@tu-dortmund.de}
\emailAdd{svjetlana.fajfer@ijs.si}

\abstract{We consider experimental  constraints  in the MeV region
in order to  determine the parameter space for the $U(1)_X$ extension of the Standard Model, presented in the first part  of our work. In particular, we focus on the model UV-completed by cold WIMPs. We conclude that the electron anomalous magnetic moment, the neutrino trident production and the relic abundance $\Omega_{CDM}$ provide the most stringent bounds and, in particular cases, they are sufficient to exclude dark-photon ($A'$) models. By allowing the axial-vector coupling of the dark gauge boson $Z'$, the interference effect with the SM  gauge bosons may reduce the bounds  coming from the trident neutrino production. At the same time,  such coupling  allows a region of the parameter space already favored both by  the relic abundance and by the discrepancy between experimental result and theoretical prediction for the muon anomalous magnetic moment.  We emphasize that light-$Z'$ interactions, non-universal for the two first lepton families, necessarily create  a difference in the proton charge radius measured in the Lamb shift of the $e$-hydrogen and $\mu$-hydrogen. Finally, we determine the effects of the new gauge boson on the forward-backward asymmetry in $e^+ e^- \rightarrow \bar{f} f$, $f = \mu, \tau$, and on  the leptonic decays $M \rightarrow j \nu_j l^+ l^-$, where $M = \pi, K, D, D_s, B$ and $j,l = e, \mu$. 
}

\begin{document} 
\maketitle
\flushbottom

\section{Introduction}
\label{sec:intro}

Phenomenology of  $U(1)$ gauge bosons $X_\mu$ (see e.g. \cite{Correia:2016xcs,Queiroz:2014zfa,Altmannshofer:2014pba,Babu:2017olk,Feng:2016ysn,Batell:2011qq,Pospelov:2008zw,Fox:2011fx}) is, in general, very dependent on the particle content and the X-hypercharge assignment of the fundamental theory. The canonical requirements for the formulation of an ultraviolet (UV) model, such as to be anomaly free and to recover the Standard Model (SM) fermion mass matrices, indicates the presence of new scalars and stable fermions even in minimal extensions like the Two Higgs Doublet Model (2HDM). Furthermore, in order to cancel the triangle anomalies per generation, it is common to introduce new right-handed fermions. Motivated by the existence of discrepancies related to the muonic interactions, by charging the second generation under $U(1)_X$, we have found an appropriate theoretical framework to discuss either dark photons or $Z^\prime$ gauge bosons. The model must contain at least two Higgs doublets and one scalar singlet, as a condition to recover a consistent fermion mass spectrum.   
Apart from its simplicity, the $U(1)_X$ SM extension is commonly found in different models as the first step of a gauge breaking scheme, like, for instance, in Grand Unified Theories  $E_6$, where $E_6 \rightarrow SO(10) \otimes U(1) \rightarrow SU(5) \otimes U(1) \otimes U(1) \rightarrow SM \otimes U(1)$.

Guided by our theoretical analysis of $\text{SM}\otimes U(1)_X$ theories, in this paper we describe the constraints from the existing experimental data  on the  dark gauge boson  phenomenology at  the low energies  (MeV regime). We first consider leptonic interactions and provide the detailed expressions for the computation of the  most stringent processes. In Part I of our work we present how the set of $X_\mu$ gauge bosons can be separated into the dark photons $A'$ and the neutral gauge bosons $Z'$ subsets, depending whether they have axial couplings with fermions or not. In the minimal dark photon phenomenology \cite{Alexander:2016aln}, only three new parameters are present, namely, the kinetic-mixing coupling, the mass of dark photon  $A'$  and its branching fraction into invisibles. In the more general case when SM fields are charged under $X$, the parameter space is increased by at least two parameters - the gauge coupling, and the new breaking scales. The  presence of dark photon  leads to small corrections of  Quantum Electrodynamic (QED) quantities 
and is commonly favored in current experimental searches, for falsifiability reasons and safety from parity-violating effects. Notwithstanding, we aim to  explicitly obtain the dependence of axial couplings on the model parameters and  why their effects must be suppressed. We also check  what is the effect of axial couplings  in  the decay width of $Z'$ into  dark matter candidates. In  the chiral $U(1)_X$ model, Lepton Flavor Violation  (LFV) is  directly suppressed 
through  the presence of a flavor matrix $\mathbb{F}$.  For instance, by X-charging the second generation of the right-handed fermions and selecting the element $\mathbb{F}_{22} = 1$, we immediately achieve  that no LFV emerges at tree-level.

The analysis of Sec.\ref{Sec:Const} is performed under the hypothesis that there is a light $X_\mu$ boson below the di-muon threshold ($\sim 10 - 200$ MeV),  whose effects can be detected for couplings of the order $g_X \sim 10^{-4}-10^{-1}$. If this hypothesis holds, the gauge boson existence must be associated  to one specific point of the most favored (less excluded) parameter space, in the  plane defined by $m_X$ and $g_X^2$. 
We will see that the search for the most favored parameter planes will imply, for instance, that the fermionic dark matter candidate $\chi$ is lighter than the new vector, once the decaying channel $X \rightarrow \bar{\chi} \chi$ is open. In addition, the strong bounds from the electron anomalous magnetic moment \cite{Pospelov:2008zw} can be reduced in the vicinity of the poles, which emerge in the calculations  of the $(g-2)_e$,  provided  by the presence of axial couplings. This feature might totally exclude dark photon models coupled to dark matter fermionic candidates.  A similar conclusion is reached  from  the neutrino trident production  \cite{Altmannshofer:2014pba}. 

We denote by $X_\mu$  a generic new boson, while $Z'$ and $A'$ are assigned to a new neutral gauge  and dark photon, respectively, whenever the distinction is necessary in our analysis. In Sec.\ref{Sec:Out} we calculate the forward-backward asymmetry for  $\bar{e} e \rightarrow \bar{f}f$, which is relevant in    the studies of light $Z'$ physics.  
In Sec.\ref{Sec:Mmu2ee} we make predictions for leptonic meson decay widths for $M \rightarrow j \nu_{j} \bar{l}l$, with $M = \pi, K, D, D_s B$ and $j,l = e, \mu$, in the framework of $SM \otimes U(1)_X$. Finally, Sec.\ref{Sec.Coc} contains a summary of our  results.

\section{ \texorpdfstring{$U(1)_X$}{} Coupling to Right-Handed Fermions}
The present work considers a  number of the most stringent constraints coming from the  flavor physics in the MeV regime, applied for a particular $U(1)_X$ SM extension which exclusively charges  the second generation of all right-handed (RH) fermion species. Here we summarize the important vertexes, necessary for  the computation presented in the next section. The fermion-gauge interactions are described by:
\begin{equation}\label{Eq.GI}
\mathcal{L}_{kin} \supset i \biggl[ \overline{L}_{\alpha L} \slashed{D} {L}_{\alpha L} 
+
\overline{Q}_{\alpha L} \slashed{D} {Q}_{\alpha L} 
+ 
\overline{l}_{\alpha R} \slashed{D} {l}_{\alpha R} 
+ 
\overline{d}_{\alpha R} \slashed{D} {d}_{\alpha R} 
+ 
\overline{u}_{\alpha R} \slashed{D} {u}_{\alpha R} 
+ 
\overline{\chi}_{ R} \slashed{D} {\chi}_{ R} \biggr], 
\end{equation}
where $\alpha = 1,2,3$, $L, Q$ are lepton and quark isospin doublets, while $l,u,d$ are the lepton, d-type and u-type fermion singlets, respectively.  The covariant derivative $D_\mu$ is introduced as
\begin{equation}\label{Eq.Cov}
D_\mu = \partial_\mu - i g(W^+ \mathbb{I}_+ + W^- \mathbb{I}_-) -i e \mathbb{Q} A_\mu -i  g_Z Z_\mu
-i g_R X_\mu .
\end{equation}
The new $SU(2)$ singlet and stable fermion,  $\chi_R$, is required for the treatment of quantum anomalies. The couplings in the mass basis are dependent on two independent angles, i.e.
\begin{subequations}\label{Eq.coup2}
	\begin{eqnarray}
	\label{couplings1}
e \mathbb{Q} &=&  g s_\phi \tau^3 + g_Y c_\phi Y , \\
\label{couplings2}
g_Z &=& c_\theta g_Z^{SM} + s_\theta (\kappa Y + g_X X), 
\\
\label{couplings3}
g_R &=& s_\theta g_Z^{SM} - c_\theta (\kappa Y + g_X X),
\end{eqnarray}
\end{subequations}
where $g, g_Y, g_X$ are the weak couplings and $\tau_3$ is the $SU(2)$ generator.  We use $c_\phi\equiv  cos \phi$ and $s_\phi \equiv sin \phi$. The quantum numbers $Y$ and $X$ are related to the $U(1)_Y$ and $U(1)_X$  parts of the $\text{SM}\otimes U(1)_X$ gauge group. The  parameter $\kappa = g_Y \epsilon $ is the coupling resulting from the ${\cal L} _m=\frac{\epsilon}{2} B^Y_{\mu \nu}B^{X \mu \nu}$ kinetic mixing term, while $g_Z^{SM} = \frac{g}{c_\phi} (\tau_3 -s^2_\phi \mathbb{Q})$. There are two Higgs doublets and a singlet in the model, denoted by:
\begin{equation}
\phi_0 = 
\begin{pmatrix}
\varphi_0^+ \\ \frac{v_0 + H_0 + i \chi_0}{\sqrt{2}}
\end{pmatrix}, \qquad
\phi_X = 
\begin{pmatrix}
\varphi_X^+ \\ \frac{v_X + H_X + i \chi_X}{\sqrt{2}}
\end{pmatrix}, \qquad
s = \frac{v_s + H_s + i \chi_s}{\sqrt{2}}.
\end{equation}
with the following hypercharges
\begin{equation}
Y_{0} = Y_{X} = \frac{1}{2}, \quad X_{0} = 0, \quad X_{X} = -1, \quad 
Y_s = 0 \quad X_s = 1\,.
\end{equation}
The vacuum expectation values $v_0, v_X$ are the weak breaking scales assumed to be related to the SM $v^2 = v_0^2 + v_X^2$. The full reproduction of QED and the introduction of a new breaking scale, larger than the SM weak scale, allow the angles to be parametrized as:
\begin{equation}\label{Eq.steta}
s_\phi = \frac{g_Y}{\sqrt{g^2 + g_Y^2}}, \qquad s_\theta \approx \frac{|2 g_X c^2_\beta - \kappa|}{\bar{g}} \left[1 - \frac{m^2_X}{m^2_Z}\right]^{-1}.
\end{equation} 
and $c_\beta^2 = \frac{v_X^2}{v^2}$. The parameter $s_\theta$ regulates the NP effects in the neutral currents and it must be a small parameter. The $g_X X$ term of Eq.(\ref{Eq.coup2}) operates for RH fields and generates a so-called flavor matrix $\mathbb{F}$, encompassing the flavor changing (FCNC) and non-universality effects in the theory. $\mathbb{F}$ emerges after the field redefinition $f_R \rightarrow V_{fR} f_R' \equiv  V_{fR} f_R$, where $f = (f_1, f_2, f_3)$:
\begin{eqnarray}\label{Eq.GIb}
\mathcal{L}_{kin} &\supset& - c_\theta g_X \biggl[ \overline{u}_{R} \mathbb{F}^U \gamma^\mu {u}_{R} 
\ + \overline{d}_{R} \mathbb{F}^D \gamma^\mu {d}_{R} 
\ + \overline{l}_{R} \mathbb{F}^l \gamma^\mu {l}_{R}
\biggr] X_\mu, 
\end{eqnarray}
such that $
(\mathbb{F}^f)_{ij} = X^f (V^\dagger_{fR})_{i2} (V_{fR})_{2j}$, with $
\text{Tr}[\mathbb{F}^f]
= \text{Tr}[\mathbb{X}^f] 
=X^f$, where $X^f$ is the X-hypercharge of the particular fermion type. From the definition
\begin{equation}
|\mathbb{F}^f| \equiv X^f \begin{pmatrix}
|V_{f R}|^2_{21} & |V_{fR}|_{21} |V_{fR}|_{22} & |V_{fR}|_{21} |V_{fR}|_{23} \\ |V_{fR}|_{21} |V_{fR}|_{22} & |V_{f R}|^2_{22} & |V_{fR}|_{22} |V_{fR}|_{23} \\ |V_{fR}|_{21} |V_{fR}|_{23} & |V_{fR}|_{22} |V_{fR}|_{23} & |V_{f R}|^2_{23}
\end{pmatrix}\,, 
\end{equation}
one can verify that the $\mathbb{F}$ diagonal elements control the amount of Lepton Flavor Violation (LFV) predicted by the model. 

In summary, the New Physics (NP) effects in flavor are dominated by $X_\mu$ interactions, which can be represented by\footnote{NP in $Z$ interactions are doubled suppressed by $s_\theta g_X \approx g_X^2$.}
\begin{equation}\label{Eq.NeutralCur}
\mathcal{L} \supset \frac{1}{2} \ \overline{f} \ \gamma_\mu (x_V^f + x_A^f \gamma^5) \ f \ X^\mu. 
\end{equation}
The couplings $x_{V,A}^f$ can be extracted by replacing in Eq.(\ref{Eq.Cov}) and Eq.(\ref{Eq.GI}) the following hypercharge assignment:
\begin{itemize}
	\item $Y$ hypercharges: \begin{equation}
	Y_L = -\frac{1}{2}, \quad Y_Q = \frac{1}{6}, \quad Y_l = -1, \quad Y_\chi = 0, \quad Y_u = \frac{2}{3}, \quad Y_d = - \frac{1}{3}\,,
	\end{equation}
	\item $X$ hypercharges (the second generation receives the charges under $U(1)_X$): 
	\begin{equation}
	X_L = 0, \quad X_Q = 0, \quad X_{e2} = 1, \quad X_{\chi_R} = -1, \quad X_{u2} = -1, \quad X_{d2} = 1\,,
	\end{equation}
	where the remaining RH fields uncharged.
\end{itemize}
For completeness, we present the quarks and leptons vector and axial-vector couplings 
\begin{subequations}\label{Eq.coup}
	\begin{eqnarray}
	x_V^U &=&  g \frac{s_\theta}{c_\phi} \left(\frac{1}{2} - \frac{4}{3} s^2_\phi \right)
	- c_\theta \kappa \frac{5}{6} - c_\theta g_X \mathbb{F}^U_{ii},
	\\
	x_A^U &=&  g \frac{s_\theta}{c_\phi} \left(- \frac{1}{2} \right)
	- c_\theta \kappa \frac{1}{2} - c_\theta g_X \mathbb{F}^U_{ii},
	\\
	x_V^D &=&  g \frac{s_\theta}{c_\phi} \left(- \frac{1}{2} + \frac{2}{3} s^2_\phi \right)
	+ c_\theta \kappa \frac{1}{6} - c_\theta g_X \mathbb{F}^D_{ii},
	\\
	x_A^D &=&  g \frac{s_\theta}{c_\phi} \left(\frac{1}{2}\right)
	+ c_\theta \kappa \frac{1}{2} - c_\theta g_X \mathbb{F}^D_{ii},
	\\
	x_V^l &=&  g \frac{s_\theta}{c_\phi} \left(- \frac{1}{2} + 2 s^2_\phi \right)
	+ c_\theta \kappa \frac{3}{2} - c_\theta g_X \mathbb{F}^l_{ii} \label{Eq.xvl},
	\\
	x_A^l &=& g \frac{s_\theta}{c_\phi} \left(\frac{1}{2}\right)
	+ c_\theta \kappa \frac{1}{2} - c_\theta g_X \mathbb{F}^l_{ii},
	\\
	x_V^\nu &=&  - x_A^\nu = g \frac{s_\theta}{c_\phi}  \left(\frac{1}{2}\right)
	+ c_\theta \kappa , \label{Eq.xvnu}
	\\
	x_V^\chi &=& x_A^\chi = c_\theta g_X.
	\end{eqnarray}
\end{subequations}
We emphasize that the effects of additional new scalar fields are considered to be negligible, in contrast with those  coming from the $\chi$ fermion coupling to $X_\mu$.

\paragraph{\texorpdfstring{$X_\mu$ interactions with charged hadrons}{}}
 The interaction of a new dark gauge boson $X_\mu$ with charged hadrons can be obtained by using the gauge principle of  the QED 
Lagrangian, and by rotating the Abelian gauge field like 
\begin{equation}\label{Eq:Red1}
B^Y_{\mu} \rightarrow B^Y_{\mu} + \epsilon B^X_{\mu}, 
\end{equation}
which brings the kinetic mixing, at first order in $\epsilon$, into a diagonal form. In other words, after the above transformation, the QED covariant derivative $
D_\mu = \partial_\mu -i e q A_\mu
$ is extended  to 
\begin{eqnarray}\label{Eq.Had}
D_\mu \stackrel{(\ref{Eq:Red1})}{\rightarrow} \partial_\mu -i e q A_\mu + i q c^2_\phi \kappa X_\mu, 
\end{eqnarray}
where $\kappa = g_Y \epsilon$.  We neglect second order terms in the small parameters. The shift will allow us to compute the inner $X$-bremsstrahlung from a charged hadron. We present in Sec.\ref{Sec:Mmu2ee} its importance in the case of  the  $M\to \mu \nu ee$ meson decays, for $M = \pi, K, D, D_s$ and $B$.

\section{Low-Energy Constraints}\label{Sec:Const}

\subsection{\texorpdfstring{$\rho$}{} Parameter} 
The dependence of the $W$-boson mass  on the coupling and vacuum expectation values (v.e.v) reproduces the Standard Model value  at tree-level, i.e.
\begin{eqnarray}\label{eq.WZmass}
m^2_W = \frac{g^2 v^2}{4} , \enspace \text{with} \enspace v^2 \equiv v_X^2 + v_0^2.
\end{eqnarray}
In the limit of vanishing new couplings, the $Z$-boson mass is SM-like,  while   the mass of $X_\mu$ gauge boson becomes 
$M_X^2\to g_X^2 \bar v^2 +\kappa^2 v^2/4$  (with $\bar v^2 \equiv v_X^2 + v_s^2$, defined in Eq. (2.21) of the Part I).
The $\rho$ parameter is defined by three observables, namely $m_W$, $m_Z$ and the weak mixing angle, through the expression
$\rho = \frac{m^2_W}{m^2_Z c^2_w}$. We use notation $cos \theta_w \equiv c_w$. In the SM these parameters are connected, such that  at tree-level $\rho = 1$. Within $\text{SM}\otimes U(1)_X$ theories, if the couplings $g$ and $g_Y$ are assumed to take the SM values, or equivalently $c_\phi \equiv c_w $, the Z mass parameter approaches $(m^2_{Z})_{SM}$ from the right, i.e. $(m^2_{Z})_{X} > (m^2_{Z})_{SM}$, which leads to  a suppression in $\rho$.  In order to find how the $\rho$ parameter differs from unity, we can write $m_Z$ as
\begin{equation}
m^2_Z \approx \frac{v^2}{4} \bar{g}^2 \left(1 + \frac{a_2^2}{\bar{g}^2 - a_1}\right) \rightarrow \frac{v^2}{4} \bar{g}^2 \left(1 + s_\theta^2\right), 
\end{equation}
with $a_1 \equiv 4 \left[ g_X^2 \frac{\bar{v}^2}{v^2} - g_X \kappa c^2_\beta \right] + \kappa^2 \,,
a_2 \equiv 2 g_X c^2_\beta - \kappa \,
$. The last step recalls Eq.(\ref{Eq.steta}) and the light mass condition $a_1 \ll \bar{g}^2$. By relying on  the above result and using the Eq.(\ref{eq.WZmass}) in the definition of $\rho$ , it follows that
\begin{equation}
\rho_X^{tree} \approx c_\theta^2 , 
\end{equation} 
which cannot reach the central value of the experimental measurement \cite{PDG:2016xqp}
$
\rho \in 1.00040(24)
$. 
Nevertheless, at two sigma level one can demand $0.99992 < c_\theta^2 \leq 1$, i.e. 
\begin{equation}
s_\theta^2 < 8 \times10^{-5}. 
\end{equation}
Hence, it follows the genuine  smallness of the $\theta$ angle at tree-level. 

\subsection{Proton Charge Radius in the \texorpdfstring{$U(1)_X$}{} Model}

The proton radius can be extracted from the  comparison of   the theoretical prediction  and the measured value for the Lamb shift in muoninc and atomic hydrogen. The result  can be expressed as a sum of independent physical contributions, i.e.
\begin{equation}
\Delta E|^l_{th} = \delta E_a^l + \delta E_b^l + \cdots + \lambda^l \langle r_p^2 \rangle |_{l}, 
\end{equation}
where $l = \mu, e$ accounts for the two types of hydrogen and the last term corresponds the correction due to the finite-size of charge distribution in the proton. At leading order $\lambda^l$ is given by
\begin{equation}
\lambda^l = \frac{2 \alpha}{3 a_l^3 n^3} \left(\delta_{P0} - \delta_{S0}\right), 
\end{equation}
where $n = 2$ for $2P -2S$ and $a_l = (\alpha m_{lp})^{-1}$ is the Bohr radius of the system with reduced mass $m_{lp} = \frac{m_l m_p}{m_l + m_p}$. Numerically  $
\lambda^\mu = - 5.2012 \ \ \text{MeV}\, \text{fm}^{-2}$.  The proton charge radius is derived from the condition  
\begin{equation}\label{Eq.proton}
\Delta {E}|^l_{th} = \Delta {E}|^l_{exp}, 
\end{equation}
with the r.h.s. denoting the experimental value of the Lamb shift in the l-hydrogen. For instance, in the muonic hydrogen \cite{Antognini:2013jkc}
\begin{equation}
\Delta E|^\mu_{exp} = 202.3706(23) \ \text{MeV}. 
\end{equation}
On the theoretical  side, the proton charge distribution is considered to affect the effective potential defining the $\mu$-hydrogen states, whose Lamb shift is estimated to be  (see \cite{Antognini:2013jkc})
\begin{equation}
\Delta {E}|^\mu_{th} = 206.0336(15) + 0.0332(20) - 5.2275 \langle r_p^2 \rangle  . 
\end{equation}
Here the first term summarizes the vacuum polarization contributions and recoil effects, while the second includes a two-photon exchange contribution. The proton puzzle denotes the difference in the solutions of Eq.(\ref{Eq.proton}) for both $eH$ and $\mu H$ systems, which provides\footnote{Note that the level of precision in the $\mu H$ is one order of magnitude higher than in $eH$.} 
\begin{subequations}\label{Eq.Pr0}
	\begin{eqnarray}
\sqrt{\langle r_p^2 \rangle}|_{\mu} &=& 
0.84087(39) \ \text{fm}      \qquad \text{\cite{Antognini:2013jkc}}
\\
\sqrt{\langle r_p^2 \rangle}|_{e} &=& 
0.8758(77) \ \text{fm}  \qquad \text{CODATA-2010 \cite{Codata2010}}
\end{eqnarray}    
\end{subequations}
The above results are obtained when no New Physics effects are included. 
In the present work we accommodate this discrepancy through the  NP contribution  in Eq.(\ref{Eq.proton}), whose l.h.s. may  be rewritten as
\begin{equation}\label{Eq.PLamb}
\Delta E|^l_{th} = \delta E_0^l + \delta E_X^l +  \lambda^l \langle r_p^2 \rangle |^X_{l}
\end{equation}
where $\delta E_0^l$ sums up the errors of the results in Eq.(\ref{Eq.Pr0}), and now
\begin{equation}\label{Eq.Pbound}
\langle r_p^2 \rangle |^X_{\mu} = \langle r_p^2 \rangle |^X_{e}. 
\end{equation}
Moreover, from Eq.(\ref{Eq.PLamb}) the difference between the "X" and QED frameworks can be expressed as a small deviation $\delta_l^X$:
\begin{equation}
\langle r_p^2 \rangle|_{l}^X  = 
 \langle r_p^2 \rangle|_{l} - \delta_l^X, 
\end{equation}
with
\begin{equation}\label{Eq.deltaproton}
\delta_l^X \equiv \frac{\delta E_X^l}{\lambda^l}. 
\end{equation}
In summary, a constraint on the proton radius  is imposed by Eq.(\ref{Eq.Pbound}), i.e.
\begin{equation}
\delta_e^X - \delta_\mu^X = \langle r_p^2 \rangle|_{e} - \langle r_p^2 \rangle|_{\mu}. 
\end{equation}
In general, the correction $\delta_l^X$ is derived as the deviation from the Coulomb potential due to the exchange of a massive vector boson $X_\mu$, or (\cite{TuckerSmith:2010ra})
\begin{equation}
V^l_X (r) = \frac{g_l g_p}{e^2} \frac{\alpha e^{- m_X r}}{r}, 
\end{equation}
with a correspondent shift in $2P-2S$ given by \cite{Pachucki:1996zza}
\begin{eqnarray}
\delta E_X^l &=& \int dr \ V^l_X(r) \left(|R_{21}(r)|^2 - |R_{20}(r)|^2\right) r^2 
= - \frac{\alpha}{2 a_l^3} \left(\frac{g_l g_p}{e^2}\right) \frac{f(a_l m_X)}{m^2_X}.
\end{eqnarray}
Above, $|R_{2i}|^2$ are the radial wave-functions corresponding to the states $P$ and $S$, while  $g_l, g_p$ are the lepton and proton couplings with $X_\mu$, respectively and  $f(x) = \frac{x^4}{(1+x)^4}$ \cite{TuckerSmith:2010ra}. The parameter $a_l = (\alpha m_{lp})^{-1}$ denotes the Bohr radius, with $m_{lp}$ denoting  the reduced mass of the $l-p$ system, and implies $a_\mu \sim 1.44$, i.e. $\sim \frac{m_\mu}{m_e}$ smaller than $a_e$. Therefore, for $m_X > 10$ MeV one can  approximate $f(x)  \sim 1$. From Eq.(\ref{Eq.deltaproton}) it follows that
\begin{equation}\label{Eq.deltaprotonII}
\delta_l^X = 6 \left(\frac{g_l g_p}{e^2}\right) \frac{f(a_l m_X)}{m^2_X}
\end{equation}
and a \textit{proton curve} is defined by
\begin{equation}
6 \frac{g_p}{e^2} \frac{(g_e - g_\mu)}{m^2_X} = \langle r_p^2 \rangle|_{e} - \langle r_p^2 \rangle|_{\mu}, 
\end{equation}
which, in principle, can be solved by an attractive force (i.e. $ \text{sgn}g_p = - \text{sgn}g_l$) strongly coupled with muons. 
In the $\text{SM}\otimes U(1)_X$ framework, and in the limit $f(x) \sim 1$, the $\text{sgn} \ g_p$ must be opposite only to the non-universal part of the $X^\mu$ coupling. The couplings $g_p$ and $g_l$, extracted from Eq.(\ref{Eq.Had}) and Eq.(\ref{Eq.xvl}) respectively, are given by:
\begin{equation}
	g_p = - c_\phi^2 \kappa, \qquad g_l = \frac{x_V^l}{2}.
\end{equation}
For simplicity $\mathbb{F}_{\tau \tau}$ may be set to 0, such that $\mathbb{F}_{\mu \mu} + \mathbb{F}_{ee} = 1$, what reduces the proton curve to
\begin{equation}
6 \frac{g_p g_X}{e^2} \frac{2\mathbb{F}_{\mu \mu} - 1}{m^2_X} = 0.060(13) \ \text{fm}^2.
\end{equation}


In  Fig.\ref{fig:Pang2} we present  two examples for the  $2 \sigma$ favored region allowed by the current experimental results  for the proton radius\footnote{Note that since $g_p = - c_\phi^2 \kappa$, a force strongly coupled to electrons could also explain the anomaly through an opposite phase in the kinetic mixing constant.}, with the fixed parameters $(c_\beta, \kappa, \mathbb{F}_{\mu\mu}) = (0.8, -4g_X, 1)$ and $(c_\beta, \kappa, \mathbb{F}_{\mu\mu}) = (0.8, -g_X, 1)$. We emphasize that,  even in the absence of the experimental discrepancy  in Eq.(\ref{Eq.Pr0}), any theory of lepton non-universality in  the first two generations will imply, from Eq.(\ref{Eq.PLamb}), a non-zero contribution to the Lamb shift in the $l$-hydrogen system and, therefore, it will impose strong constraints on  these couplings in the MeV range.

\subsection{\texorpdfstring{$M^+ \rightarrow \mu^+ $ invisibles  }{}}
In this section we compare the experimentally measured  decay width for $K \rightarrow \mu Y$ ($Y$ denotes invisible states), 
 and  the theoretical prediction  for the decay width of $K \rightarrow \mu^+ \nu_\mu \chi \bar{\chi}$. The narrow-width approximation (NWA) is assumed to be valid in the region where $m_X > 2 m_\chi$, i.e when $e^+ e^-$, $\nu \overline{\nu}$ and $\chi \bar{\chi}$ are the only directly accessible decay products of $X_\mu$. Hence the differential decay width may be determined  using 
 \begin{equation}\label{Eq.Decay}
d\Gamma (K \rightarrow \mu \nu \chi \bar{\chi}) = \frac{1}{3} d\Gamma (K \rightarrow \mu \nu X) \text{Br}(X \rightarrow \chi \bar{\chi}). 
\end{equation} 
The dominant contributions to the decay amplitude are presented in  Fig.\ref{fig:KmuX}(a,b). The Feynman's rules for the vertexes related to Fig.\ref{fig:KmuX}(a,b) are given in Appendix \ref{Ap.FR} and they lead  to  the amplitudes
\begin{subequations}
\begin{eqnarray}
\mathcal{M}_a & = &  \left(\frac{G_F}{2 \sqrt{2}} f_M V^*_{UD}\right) \frac{\epsilon^{\mu *}_X}{q_{23}^2 - m^2_l} \Biggl\{(x_V^l + x_A^l)m_l \left[ \bar{u}_\nu (1 + \gamma_5) \left((\slashed p_2 + m_l)\gamma_\mu + 2p_{3\mu} \right) v_l\right] - \nonumber \\
& & - (x_V^l - x_A^l) q_{23}^2 \left[\bar{u}_\nu (1 + \gamma_5) \gamma_\mu v_l\right] \Biggr\} \\
\mathcal{M}_b & = & \left(\frac{G_F}{\sqrt{2}} f_M V^*_{UD}\right) (c_\phi^2 \kappa) 
\left[(k + q)\cdot \epsilon^*_X\right] \frac{m_l}{q^2 - M^2} \left[\bar{u}_\nu (1 + \gamma_5) v_l\right],
\end{eqnarray}
\end{subequations}
where 
$q = p_ 1 + p_3$ and $q_{23} = p_2 + p_3$, $f_M$ are meson decay constants and $V_{UD}$ stands for the CKM matrix element present in the particular decay mode. 
As a cross-check one can prove that the gauge invariance holds by replacing   the dark gauge boson $X_\mu$ by the photon field through $x_V^l \rightarrow 2e, x_A^l \rightarrow 0$ and $c_\phi^2 \kappa \rightarrow e$, such that  $\epsilon_\mu (\mathcal{M}_a^\mu + \mathcal{M}_b^\mu) \rightarrow p_{2\mu} (\mathcal{M}_a^\mu + \mathcal{M}_b^\mu) = 0$. 

\begin{figure}%
	\centering
	\subfloat[]{{\includegraphics[width=4.5cm]{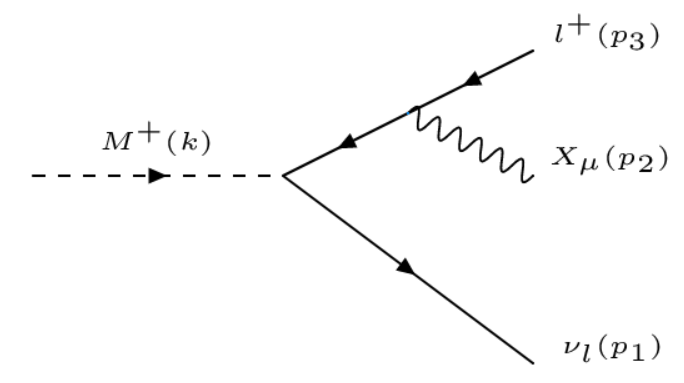} }}%
	\quad	
	\subfloat[]{{\includegraphics[width=4.5cm]{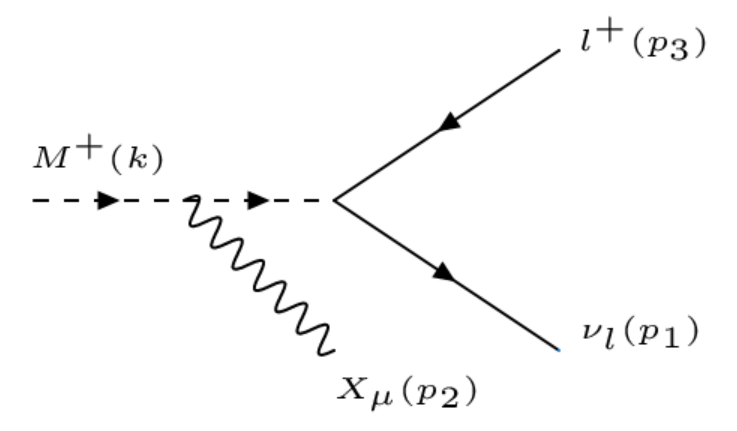} }}%
	\quad
	\subfloat[]{{\includegraphics[width=4.5cm]{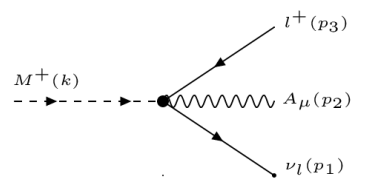} }}%
	\caption{The Feynman diagrams (a,b) contributing to $M_{l Y}$, Y denoting invisibles, in the $U(1)_X$ model and the structure dependent diagram (c) in QED.}%
	\label{fig:KmuX}%
\end{figure}
The details of the decay width calculation are given in Appendix \ref{AP.DW}. 
Assuming that the vector $X_\mu$ decays  into the invisible pairs $\bar{\chi} \chi, \bar{\nu} \nu$, its contribution to the process $K \rightarrow \mu + invisibles$ can be constrained by  the 
existing experimental bound \cite{Pang:1989ut} 
\begin{equation}\label{Eq.Pang}
\frac{\Gamma_{K \mu Y}}{\Gamma_{K \mu \nu}} < 3.5 \times 10^{-6}, \qquad 90\% \ C.L.\,, 
\end{equation}
with the missing energy in the interval 
$227.6 < m_Y (MeV) < 302.2$.  The vertex $X_\mu \bar{\chi} \chi$ is extracted from the Lagrangian component
\begin{equation}
\mathcal{L} \supset  \frac{g_R}{2} X_\mu \bar{\chi} \gamma_\mu (1 + \gamma_5) \chi , 
\label{Xchichi}
\end{equation}
with $g_R$ given by Eq.(\ref{couplings3}). Since $\chi_R$ is a singlet under the SM gauge group and $X_\chi = -1$, it follows that the Feynman's rule for the $X_\mu \bar{\chi} \chi$ 
is $i (g_X/ 2 )\gamma_\mu (1 + \gamma_5)$,  where we  set  $c_\theta \sim 1$. 
If the mass of $X_\mu$  is in  the MeV region, the dark boson may still decay into a electron-positron pair, whose vertex is written in Eq.(\ref{Eq.evert}). In the NWA,
 \begin{eqnarray}
\Gamma_{K \mu Y} &=& \Gamma_{K \mu \bar{\chi}\chi}  + 3 \Gamma_{K \mu \bar{\nu}\nu} \nonumber \\
&=& \frac{1}{3} \Gamma (K \rightarrow \mu \nu X) \left[\text{Br}(X \rightarrow \chi \bar{\chi}) + 3\text{Br}(X \rightarrow \nu \bar{\nu}) \right], 
\end{eqnarray}
where the  factor of three accounts for the neutrino flavors. 
\begin{figure}[tbp]
	\centering 
		\includegraphics[width=.35\textwidth]{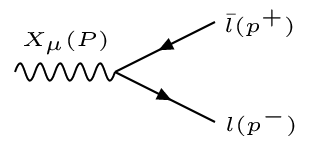}
	\caption{\label{fig:Xll} For $m_X < 2m_\mu$ the boson $X_\mu$ has five directly decay modes, $l = e, \chi, \nu$.}
\end{figure}
The decay amplitude coming from  the diagrams in  Fig.\ref{fig:Xll} can be  written  most  generally as  
\begin{equation}\label{Eq.Ml}
\mathcal{M}_l = \frac{1}{2} \left[\bar{u}(p^-) \gamma_\mu (x_V^l + x_A^l \gamma_5) v(p^+)\right] \epsilon^\mu (P). 
\end{equation}
From Eq.(\ref{Xchichi}), $x_V^\chi = x_A^\chi = g_X$. Again, one can estimate the branching ratio  
\begin{equation}
BR(X \rightarrow \bar{a}a) = \frac{\Gamma(X \rightarrow \bar{a}a)}{\sum_l \Gamma(X \rightarrow \bar{l}l)}
\end{equation}  
where, from Eq.(\ref{Eq.DecayFormula}), 
\begin{equation}
\Gamma(X \rightarrow \bar{l}l) = \frac{\sum |\mathcal{M}_{X\bar{l}l}|^2}{2 m_X} \frac{\Phi_2(X\rightarrow \bar{l}l)}{(2\pi)^2},
\end{equation}
i.e.
\begin{equation}
BR(X \rightarrow \bar{a}a) = \frac{|\mathcal{M}_{X\bar{a}a}|^2 \sqrt{\lambda(m_X^2, m_a^2, m_a^2)}}{\sum_l |\mathcal{M}_{X\bar{l}l}|^2 \sqrt{\lambda(m_X^2, m_l^2, m_l^2)}}. 
\end{equation}
The sum over $l$ takes all  $l = \chi, e, \nu_e, \nu_\mu, \nu_\tau$.
Finally, from the amplitude of Eq.(\ref{Eq.Ml}), one can write the general formula
\begin{equation}
|\mathcal{M}_{X\bar{l}l}|^2 = 4 \left[2 m_l^2 \left({x_V^{l}}^2 -2 {x_A^{l}}^2\right) + m_X^2 \left({x_V^{l}}^2 + {x_A^{l}}^2\right)\right], 
\end{equation}
where the average over final state spins is made implicit. Since $x_V^\chi = x_A^\chi$, the equation is simplified, such that  
\begin{equation}
|\mathcal{M}_{X\bar{\chi}\chi}|^2 \propto [m_X^2 - m_\chi^2 ]
\end{equation}
i.e. the decay width of $X \to \chi \bar \chi$ is suppressed by the difference between $X$ and $\chi$ masses. 
Provided by the experimental constraint \cite{Pang:1989ut},  we present some examples of the allowed region for $m_X \times g_X^2$ in Fig.\ref{fig:Pang2} and Fig.\ref{fig:g}.

\subsection{\texorpdfstring{$(g-2)_e$}{}}

The contribution  to the electron anomalous moment  $a_e$,  coming from the  new dark gauge boson $X_\mu$ is equivalent to a shift in the fine-structure constant, as already discussed in Ref. \cite{Pospelov:2008zw}
\begin{equation}
d\alpha = 2 \pi a_e^X \quad \rightarrow \quad 
\frac{d \alpha^{-1}}{\alpha^{-1}} = - \frac{2 \pi a_e^X}{\alpha}. 
\end{equation}
The r.h.s is the relative correction to the measurement of $\alpha^{-1}$ which should not exceed $0.5$ ppb \cite{Correia:2016xcs}. The contribution of the $X_\mu$ gauge boson to the electron magnetic moment in the  dipole function  can be written as 
\begin{equation}
a_e^X = \frac{m_e^2}{4 \pi^2} \left[(x_V^e)^2 I_V(m^2_X) + (x_A^e)^2 I_A(m^2_X)\right], 
\end{equation}
where
\begin{eqnarray}
I_V(m^2_X) &=& \int_0^1 dz \frac{z^2 (1-z)}{[m_l^2 z^2 + m^2_X(1-z)]} \quad \stackrel{m_X \gg m_l}{\rightarrow} \quad \frac{1}{3 m^2_X},  \nonumber \\
I_A(m^2_X) &=& \int_0^1 dz \frac{z (1-z)(z-4) - \left(2\frac{m^2_l}{m^2_X} \right)z^3 }{[m_l^2 z^2 + m^2_X(1-z)]} \quad \stackrel{m_X \gg m_l}{\rightarrow} \quad -\frac{5}{3 m^2_X}. 
\end{eqnarray}
Since the limit $m_X \gg m_e$ is valid in our analysis, we can set the bounding curve
\begin{equation}
f\left(\frac{m^2_e}{m^2_X} \right) \equiv \left(\frac{m^2_e}{m^2_X}\right) \frac{1}{6 \pi \alpha} |(x_V^e)^2 - 5 (x_A^e)^2| < 0.5 \text{ppb}
\end{equation}

\begin{figure}[tbp]
	\centering
	\subfloat[]{
		\includegraphics[width=.46\textwidth]{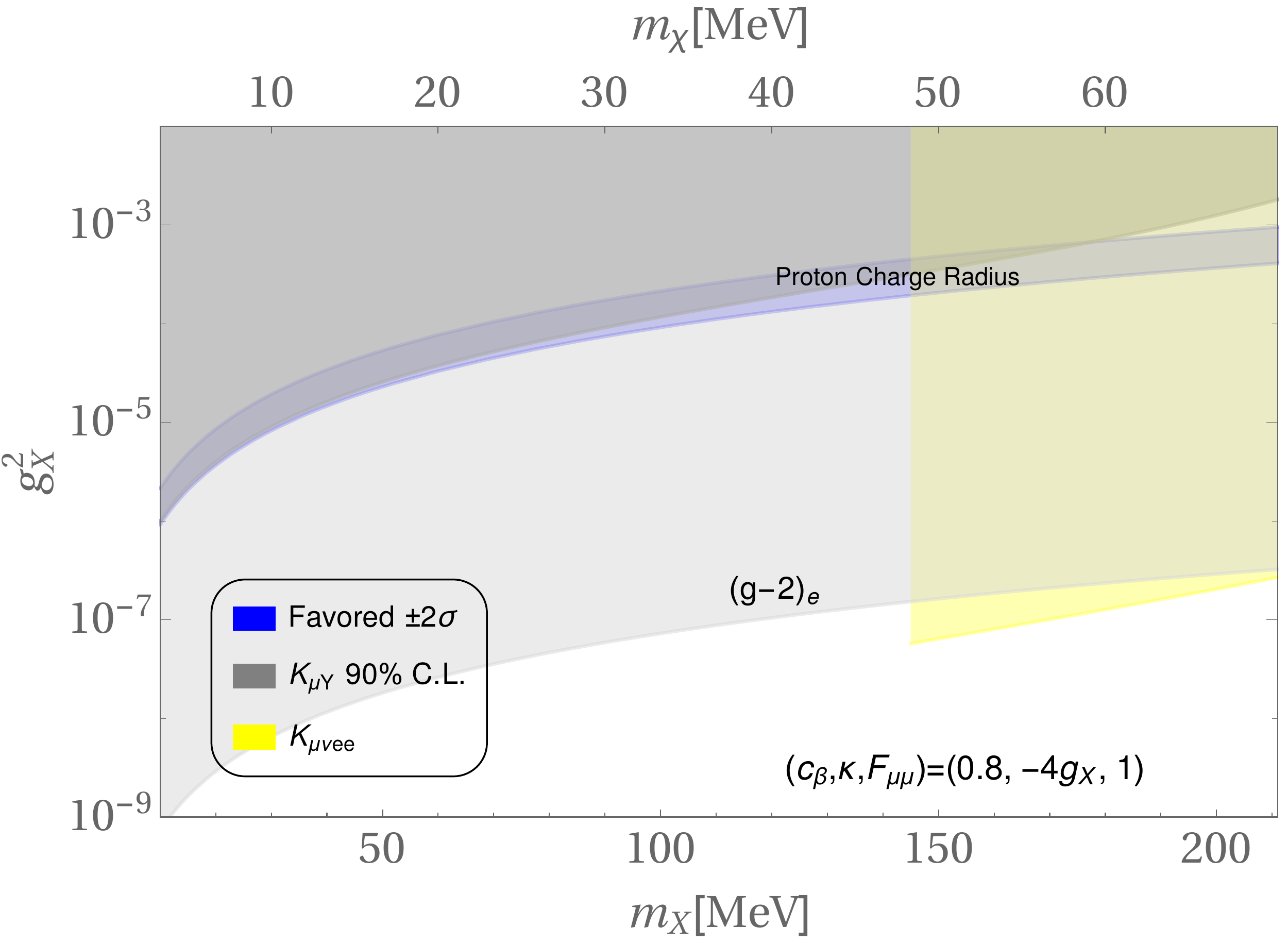}
	}
	\hspace{0.3cm}
	\subfloat[]{
		\includegraphics[width=.46\textwidth]{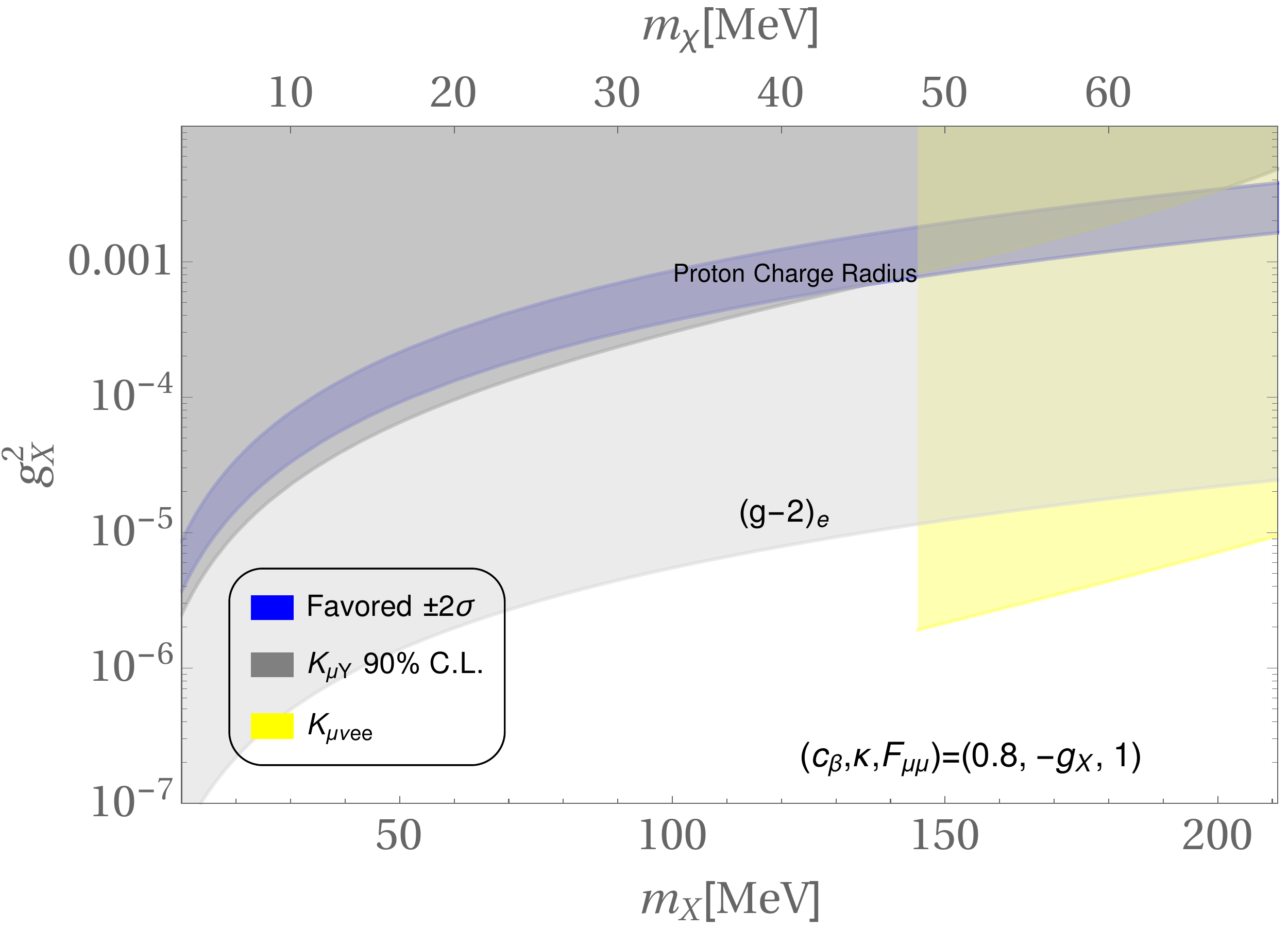}
	}
	\caption{\label{fig:Pang2} The allowed region for the proton radius  explanation,  using the bound in Eq.(\ref{Eq.Pang}). Under the narrow-width approximation the vector $X_\mu$ decays into the missing $\bar{\chi}\chi, \bar{\nu}\nu$ pairs. Here $m_X = 3m_\chi$ and $\mathbb{F}_{\tau \tau} = 0$.}
\end{figure}

\begin{figure}[tbp]
	\centering
	\subfloat[]{
		\includegraphics[width=.46\textwidth]{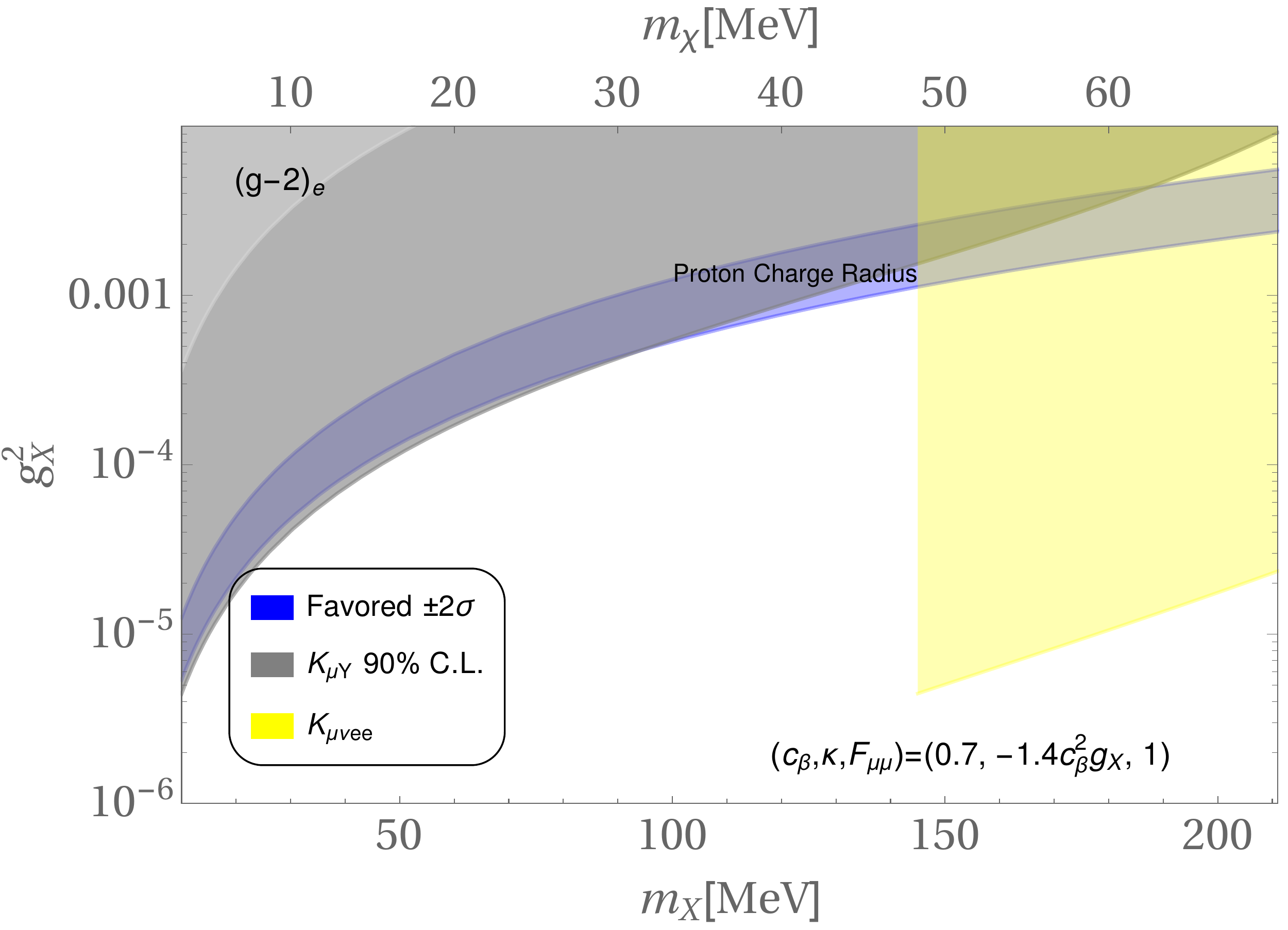}
	}
	\hspace{0.3cm}
	\subfloat[]{
		\includegraphics[width=.46\textwidth]{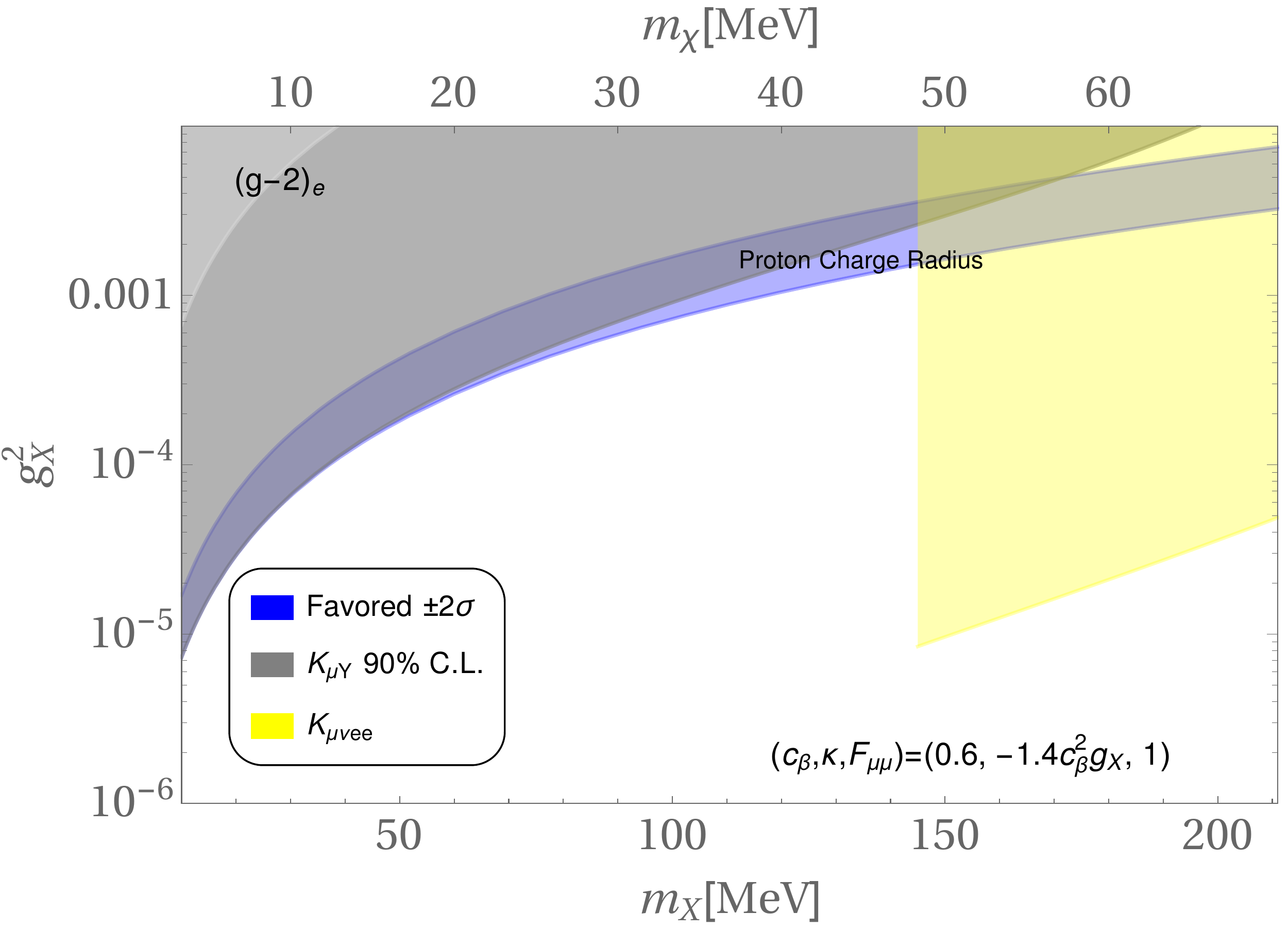}
	}
	\caption{\label{fig:g}  The allowed region for  $m_X$ and $g_X^2$ are presented, when the $(g-2)_e$ curve is minimazed by requiring the cancellation of the vector and axial contributions, as described in the text.  Two values of $c_\beta^2$ are used. The blue region is allowed by proton anomaly. 
	}
\end{figure}

\paragraph{Parameter Space}
As discussed in Part I, we have to  find out how to fix a plane in a five-dimensional parameter space  assuming that the model can explain the selected experimental discrepancies.  If we insist to  explain  the proton charge radius puzzle, one has to require
\begin{equation}\label{Eq.pre}
\text{sgn} g_X = - \text{sgn} \kappa. 
\end{equation}
In the examples depicted in Fig.\ref{fig:Pang2}, it is evident how stringent are the bounds from $(g-2)_e$. However, due to the interplay of the contributions coming from the vector and axial-vector couplings, the curve can be minimized through the root equation 
\begin{equation}\label{Eq.root}
|(x_V^e)^2 - 5 (x_A^e)^2| = 0
\end{equation}
with a fixed $\mathbb{F}$, i.e.  around the roots there is almost no effect from the dark boson $X$ to the fine structure constant. If $\mathbb{F}_{ee} = 0$ the solutions to Eq.(\ref{Eq.root}) are
\begin{equation}
n \in \left[- \frac{7}{5}, \frac{3}{2}, 3 \right] c^2_\beta. \label{conditionae}
\end{equation}
for $\kappa = n g_X$. 
Due to the condition of   Eq.(\ref{Eq.pre})  only $n = -7/5 c_\beta^2$  may be a solution for the proton puzzle,  as presented in Fig.\ref{fig:g}. We denote $\kappa_0= -7/5 c^2_\beta g_X$. 
Note that for $\kappa_0$ value, the $(g -2)_e$ bound reduces the discrepancy of the proton puzzle from $5\sigma$ to $2\sigma$.
In the case where $\kappa$  is not inside the range of  Eq.(\ref{conditionae}), then the electron anomalous magnetic moment gives the most stringent bound on the parameter space. 

In the following  calculations we have written the electron vector and axial-vector  couplings  to $X_\mu$ in the form 
\begin{eqnarray}
x_V^e &=& g_X \left[\frac{g}{c_\phi} \left(2 s_\phi^2 - \frac{1}{2}\right) \frac{|2c^2_\beta - n|}{\bar{g}} + \frac{3 n}{2} - \mathbb{F}_{ee}\right], \\
x_A^e &=& g_X \left[\frac{g}{2 c_\phi} \frac{|2c^2_\beta - n|}{\bar{g}} + \frac{n}{2} - \mathbb{F}_{ee}\right].
\end{eqnarray}
In addition, the decay of the vector boson $X_\mu$ into neutrino pairs is allowed, with couplings 
\begin{equation}
x_V^\nu =  - x_A^\nu = \dfrac{g_X}{2}\left[\frac{g}{c_\phi} \frac{|2c^2_\beta - n|}{\bar{g}} 
+ n \right]. 
\end{equation}

\subsection{\texorpdfstring{$(g-2)_\mu$}{}}
\begin{figure}[bp]
	\centering
		\includegraphics[width=.70\textwidth]{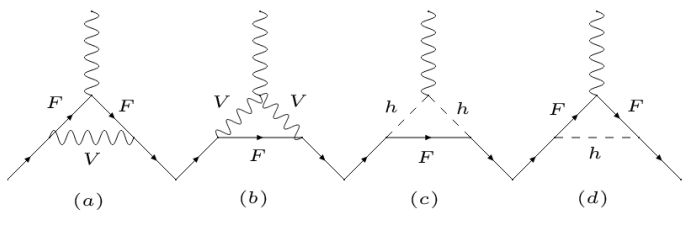}
	\caption{\label{fig:gmu} The diagrams contributing to the muon anomalous magnetic moment. The second diagram is not generated in $\text{SM}\otimes U(1)_X$ theories.}
\end{figure}

In Fig.\ref{fig:gmu} we present the most general four  contributions to the muon anomalous magnetic moment. Since the new vector boson is neutral, the diagram  (b) is the only one not contributing to our case. The diagram (c) is present for charged scalars and (d) for neutral Higgs. 
In this section we want to find the necessary conditions for the correct sign of the  $(g- 2)_\mu $ discrepancy. In the following, $h^+, h^0$ generically denote the charged and neutral scalars present in the theory. 

Following the work of \cite{Leveille:1977rc}, a general fermion-$X_\mu$ vertex can be written as
\begin{equation}
\mathcal{L} = \frac{1}{2}\sum_{F} \bar{\mu} [x_V \gamma^\rho + x_A \gamma^\rho \gamma^5] F \ X_\rho. 
\end{equation} 
For simplicity,  we do not include the suppressed flavor violating processes, i.e. $F = \mu$.
For $m_F = m_\mu$, the integral linked to the first diagram is given by
\begin{equation}
[a_\mu]_a = \frac{m^2_\mu}{16 \pi^2} \int^1_0 dz \ \frac{\left[x_V^2 [(z-z^2)z] + x_A^2[(z-z^2)(z-4) - 2\frac{m^2_\mu}{m^2_X} z^3]\right]}{m^2_\mu x^2 + m^2_X (1-x)} . 
\end{equation}
As mentioned before, we work under the assumption of very large Higgs masses, where $a_\mu$ is dominated by $[a_\mu]_a$. On the other hand, the above integral leads to  a wrong negative sign for a wide range of the $c_\beta$ parameter, and have to be compensated by additional contributions. Therefore, we assume that 
scalar masses are relatively large and compute the scalars contributions  to the moment function in the region where the asymptotic approximation to the integrals is fairly valid, i.e. $m_h > 20\, m_\mu$. The bounds on the Higgses couplings to Z are coming from the  LHC analyses,  as already considered by the authors  of Ref.   \cite{Haisch:2018kqx}. The Yukawa Lagrangian can be parametrized as 
\begin{equation}
\mathcal{L}_Y = \sum_{h,F} \bar{\mu} [C_S + C_P \gamma_5] F \ h. 
\end{equation}
Both diagrams  (c) and (d) can contribute to the muon anomalous magnetic moment. For the  diagram(c) we have to specify $F = \nu$ or $m_F = 0$. The coupling $C_P$ is in fact present in our model for both neutral and charged scalars and,  in the neutral case, it is purely imaginary. For $m_F = m_\nu = 0$ it follows that 
\begin{equation}
[a_\mu]_c = \frac{m_\mu^2}{8 \pi^2} (|C^+_S|^2 + |C^+_P|^2) \int_0^1 dz \frac{(z^3 - z^2)}{m_\mu^2 z^2 + m_{h^+}^2 (1-z)}
\end{equation}
and  for neutral scalars in the diagram (d) $F= \mu$,  one gets
\begin{equation}
[a_\mu]_d = \frac{m_\mu^2}{8 \pi^2}  \int_0^1 dz  \frac{|C^0_S|^2(2 z^2 - z^3) + |C^0_P|^2 z^3}{m_\mu^2 z^2 + m_{h^+}^2 (1-z)}, 
\end{equation}
with 
\begin{equation}
(g- 2 )_\mu = [a_\mu]_a + [a_\mu]_b + [a_\mu]_c + [a_\mu]_d. 
\end{equation}
If we consider  $m_{h^+}, m_{h^0} >> m_\mu$  the integrals converges to a simplified form: 
\begin{eqnarray}
[a_\mu]_c &\rightarrow& \frac{m_\mu^2}{8 \pi^2} (|C^+_S|^2 + |C^+_P|^2) \left(- \frac{1}{3}\right),
\\
{[a_\mu]}^S_d &\rightarrow& \frac{m_\mu^2}{m^2_{h_0}} \frac{|C^0|_S^2}{8 \pi^2} \left[\log\left[\frac{m^2_{h_0}}{m_\mu^2}\right] - \frac{7}{6}\right],
\\
{[a_\mu]}^P_d &\rightarrow& \frac{m_\mu^2}{m^2_{h_0}} \frac{|C^0|_P^2}{8 \pi^2} \left[\log\left[\frac{m^2_{h_0}}{m_\mu^2}\right] - \frac{11}{6}\right]. 
\end{eqnarray} Therefore, the charged scalars cannot  provide  the correct sign and their interactions have to be suppressed, either by their large masses or by their negligible couplings to muons.  In summary, in order to explain the  $(g-2)_\mu$ discrepancy only diagram (a) and (d) contribute. 
Again, there is a range for $c_\beta$ in which the (a) integral already gives the correct sign for the $a_\mu$ discrepancy, allowing all the scalars to live in the decoupling limit. For instance, if we do not take into account the constraint coming  from the proton radius, then we can use all solutions derived  from  Eq. (\ref{conditionae}). 
Thus, for  $\kappa = \frac{3}{2} c_\beta^2 g_X$,  and $c_\beta < 0.9$, light neutral scalars with masses in the range $m_{h^0} \in (10 - 100) m_\mu$ are required to restore $(g - 2)_\mu$, for $|C^0|_S \sim |C^0|_P \sim 10g_X$ and different values for $c_\beta$. Charged scalars are still disfavored. Since $c_\beta^2 = v_X^2/v^2$, a minimal model at low-energies (MeV) is well supported by a small scale $v_X$.
For the completeness of our study, let us mention that the pairs $\kappa = -\frac{7}{5}c_\beta^2 g_X$ and $c_\beta > 0.7$, as well as $\kappa = 3c_\beta^2 g_X$ and $c_\beta > 0.99$, both can  
solve, through (a) diagram only, the discrepancy of $(g-2)_\mu$.

\subsection{\texorpdfstring{$K_{\mu \nu e^+ e^-}$}{}}
The decay width of $K_{\mu \nu e^+ e^-}$, denoted as $\Gamma_{K\mu2ee}$,  
is obtained from  the distribution $\frac{d\Gamma_{K\mu2ee}}{dm_{ee}}$,  integrated over the electron-positron invariant  mass $m_{ee} > 145 \text{MeV}$ \cite{PDG:2016xqp}. We use  the narrow-width  approximation in a way that, for a fixed $m_X = m_{ee}$, the contribution from $X \rightarrow ee$ should not exceed the uncertainty of the total $\Gamma_{K\mu2ee}$. By demanding that the decay rate has to be smaller than the experimental uncertainty, it actually implies that no enhancement should be seen in the region $m_{ee} > 145 \,$MeV. The analysis is similar to the calculation of  $\Gamma_{K\mu Y}$, now with
\begin{equation}
\frac{1}{3} \frac{\Gamma(K \rightarrow \mu \nu_\mu X)}{\Gamma_K} \text{Br}(X \rightarrow e^+ e^-) < 3.1 \times 10^{-9}, 
\end{equation}
for $145$ MeV $< m_X < 2 m_\mu$. In Fig.\ref{fig:Pang2} and Fig.\ref{fig:g} the excluded region is marked by the yellow color. 

\subsection{Neutrino Trident Production}

\begin{figure}[tbp]
	\centering 
	\subfloat[]{
		\includegraphics[width=4.5cm]{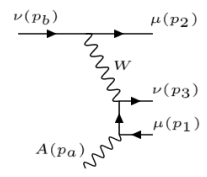}
	}
	\hspace{1.8cm}
	\subfloat[]{
		\includegraphics[width=4.8cm]{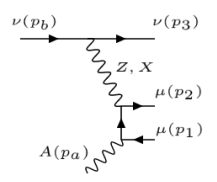}
	}
	\caption{\label{fig:NTri} The trident production in the equivalent photon approximation (EPA). In addition, there are the reciprocal diagrams where the real photon is attached to $\mu^-$.}
\end{figure}
We determine the cross-section for the neutrino trident production in the Equivalent Photon Approximation (EPA) \cite{Czyz:1964zz,Altmannshofer:2014pba,Ballett:2018uuc}, i.e. by connecting it with the scattering of a real photon and the neutrino beam. We then include the bound from the CHARM-II experiment \cite{Geiregat:1990gz}. 

The total amplitude for the scattering of a real photon and a neutrino beam in $\gamma \nu \rightarrow \nu \mu^+ \mu^-$ includes the six diagrams of Fig.\ref{fig:NTri} for the exchange of $W, Z$ and $X$ bosons. The neutrinos spins are summed,  while one takes the average over the photon polarization, i.e.
\begin{equation}
\frac{1}{2} \sum_p |\mathcal{M}|^2 = - \frac{1}{2} [\mathcal{M}]^\alpha [\mathcal{M}]^*_\alpha. 
\end{equation} 
In the case of  the neutral currents, both  $Z$ and $X$, the neutrino vector and axial-couplings are related by
\begin{equation}
x_V^\nu (z) = - x_A^\nu (z) \equiv x_z^\nu,
\end{equation}
where $z = Z, X$. The amplitude can be simplified and presented as 
\begin{equation}
[\mathcal{M}]^\alpha = \frac{G_F}{\sqrt{2}} e [\overline{u}_\nu (k_2) \gamma_\nu (1 - \gamma_5) u_\nu (k_1)]\mathcal{F}^{\alpha \nu},
\end{equation}
with
\begin{eqnarray}
\mathcal{F}^{\alpha \nu} = B_-[\overline{u}_\mu(p_-) \gamma^\alpha \{(\slashed p_- - \slashed q) + m_\mu\} \gamma^\nu (C_V + C_A \gamma_5) v_\mu(p_+)] + \nonumber \\
+ B_+[\overline{u}_\mu(p_-) \gamma^\nu (C_V + C_A \gamma_5) \{(\slashed q - \slashed p_+) + m_\mu\} \gamma^\alpha    v_\mu(p_+)]
\end{eqnarray}
and $B_i \equiv [(p_i - q)^2 - m^2_\mu]^{-1}$. 
The constants $C_{V,A}$ summarize the contributions coming from all diagrams 
\begin{equation}
C_V = \sum_{z = Z,X} \frac{x_z^\nu x^\mu_V(z)}{2 \sqrt{2} G_F} [k^2 - m^2_z]^{-1} -1, \qquad C_A = \sum_{z = Z,X} \frac{x_z^\nu x^\mu_A(z)}{2 \sqrt{2} G_F} [k^2 - m^2_z]^{-1} + 1.
\end{equation}
These expressions are exact in the sense that the Z mass parameter depends on the new $g_X, s_\theta$. Once the vertices are defined as in Eq.(\ref{Eq.NeutralCur}), by taking $c_\theta \approx 1$ and neglecting second order terms on the small parameters, the SM contribution to neutrino trident production is given by
\begin{equation}
C^{SM}_V = - \left(\frac{1}{2} + 2 s^2_\phi\right), \qquad C^{SM}_A = \frac{1}{2}. 
\end{equation}

The phase-space for the scattering of a real photon and a neutrino can be defined through the invariants \cite{Byckling:1971vca} (see Fig.\ref{fig:NTri})
\begin{equation}\label{EqNeuInv}
	s_1 = (p_1 + p_2)^2, \quad t_1 = (p_a - p_1)^2, \quad s_2 = (p_2 + p_3)^2, \quad t_2 = (p_a - p_1 - p_2)^2. 
\end{equation}
In the EPA, the  total cross-section  can be written as \cite{Belusevic:1987cw}
\begin{equation}\label{Eq.CSTrid}
\begin{aligned}
\sigma(\nu N \rightarrow \nu N \mu^+ \mu^-) &= \frac{Z^2 e^2}{64 \pi} \frac{1}{(2\pi)^5} \int^\infty _{\frac{2m_\mu^2}{E_\nu}} dq \ \int_{4m_\mu^2}^{2E_\nu q} ds \ \int_{(m_2 + m_3)^2}^{(\sqrt{s}-m_1)^2} ds_2 \int_{t_1^-}^{t_1^+} dt_1 \
\int_{t_2^-}^{t_2^+} dt_2 \\
&  
\int_0^{2\pi} d\phi
\frac{F^2(q^2) |\mathcal{M}|^2}{q s^2 \lambda^{1/2}(s,m_1^2,m_b^2)\lambda^{1/2}(s_2,m_b^2,t_1)}, 
\end{aligned}
\end{equation}
where $q \equiv |q|$, $Z$ is the atomic number of the target and $F(q^2)$ is the electromagnetic form-factor introduced in \cite{Belusevic:1987cw, Altmannshofer:2014pba}.
The details of the integration over phase space are given in Appendix \ref{AP.DW}. 
In Eq.(\ref{Eq.CSTrid}) the integrals over $q$ (with $q\equiv {\sqrt{q^2}}$) and $s = (p_a + p_b)^2$ are derived from the probability of creating a virtual photon with momenta $q$, defined by \cite{Altmannshofer:2014pba}
\begin{equation}
P(q^2,s) = \frac{Z^2 e^2}{4\pi^2} \frac{ds}{s} \frac{dq^2}{q^2} F^2(q^2) 
\end{equation}
and the electromagnetic form factor $F(q^2) = \left[1+ 1.21\frac{q^2}{m_p^2}\right]^{-2}$ is given in Ref. \cite{Czyz:1964zz}, where $m_p$ is the proton mass. In the CHARM-II experiment, a neutrino beam with the mean energy $E_\nu \sim 20$ GeV is scattered by a glass target ($Z = 10$). We require that the contribution coming from the interference 
of the SM and the $\text{SM}\otimes U(1)_X$ to the total cross-section  should be inside the one standard deviation region, i.e. 
\begin{equation}
|\sigma^{int}_{\text{\tiny SM+X}}| < 0.57\, \sigma_{\text{\tiny SM}}.
\end{equation}
For the  SM prediction, averaged over both neutrino and antineutrino scattering, we obtained 
 $\sigma_{\text{\tiny SM}} = 1.8 \times 10^{-41} \text{cm}^2$.

\subsection{ \texorpdfstring{$\chi$}{} Relic Abundance}
In this section we compute the relic abundance for the fermion $\chi$ from a set of fixed points in the parameter space. We start with the exact solution of the Boltzmann equation, aiming to examine the sensitivity of the allowed band on small variations of  the couplings. We then consider the approximate formula for the weakly interacting massive particles (WIMP)  relic density  \cite{Bertone:2010zza}
\begin{equation}\label{Eq.RAap}
\Omega h^2 \approx \frac{3 \timesß 10^{-27} \text{cm}^3 s^{-1}}{\langle \sigma_{ann} v \rangle}. 
\end{equation}
with the thermal average computed at the freeze-out temperature $T_{f.o.} \backsimeq \frac{m_\chi}{20}$.
In the general case, the attempt to create a direct bounding curve (i.e. $g_X < f(m_X^2)$ for the  analytical function $f$) is difficult due to  the presence of the $X$ boson as a resonance. In other words, the coupling $g_X$ cannot be easily factorized, once it enters the decay width in the Breit-Wigner propagator. We perform the integration  of
\begin{equation}\label{Eq.Y}
\frac{d Y}{dx} = - \left(\frac{45}{\pi M_P^2} \right)^{-1/2}\frac{g_{*}^{1/2}m_\chi}{x^2}\langle \sigma v \rangle (Y^2 - Y_{eq}^2) , 
\end{equation}
by describing the evolution of the comoving abundance Y, whose value at chemical equilibrium is given by \cite{Gondolo:1990dk}
\begin{equation}\label{Eq.Yeq}
Y_{eq} = \frac{45 g}{4 \pi^2} \frac{x^2 k_2(x)}{h_{eff}(m/x)}. 
\end{equation}
The variable $x \equiv \frac{m_\chi}{T}$, where $T$ is the photon temperature. It is commonly taken  from $x = 1$, which defines the boundary for the condition $Y = Y_{eq}$, to the present value. According to Ref.\cite{Gondolo:1990dk},  we choose for the $\chi$ mass range, $10$ MeV $ < m_\chi  < 80$ MeV, such that the effective degrees of freedom $g_{*}^{1/2}$ has a small deviation from $g_{*}^{1/2} \approx 7/2$ due to QCD quark-hadron phase transition, which we neglect in our calculation. In Eq.(\ref{Eq.Yeq}), the constant $g$ accounts the degrees of freedom for the particles present at the equilibrium. We consider the dominant channels including $\nu, e, \mu, \chi$. On the other hand, the $h_{eff}(T)$ encloses the effective degrees of freedom for entropy and, in principle, it should sum over all species present in the plasma. In practice, however, the species with large energies are suppressed by their distribution function such that we may sum over photons in addition to the particles at the equilibrium\footnote{
	The relevant species are considered to be at thermal equilibrium with the plasma and $h_{eff}(T) = \sum_i h_i(T)$ where \begin{equation}
	h_i(T) = \frac{90 g_i}{(2\pi)^5 T^3} \int \frac{3 m_i^2 + 4p_i^2}{3E_i T} \frac{p_i^2 dp_i}{\exp{E_i/T} + \eta_i}, 
	\end{equation}    
$g_i$ accounts the number of spin degrees of freedom of $i$, while $E_i, m_i$ and $\eta_i = 1 (-1)$ denote its energy, mass and Fermi-Dirac (Bose-Einstein) statistics, respectively.}. 

\begin{figure}
	\centering
	\subfloat[]{
		\includegraphics[width=10cm]{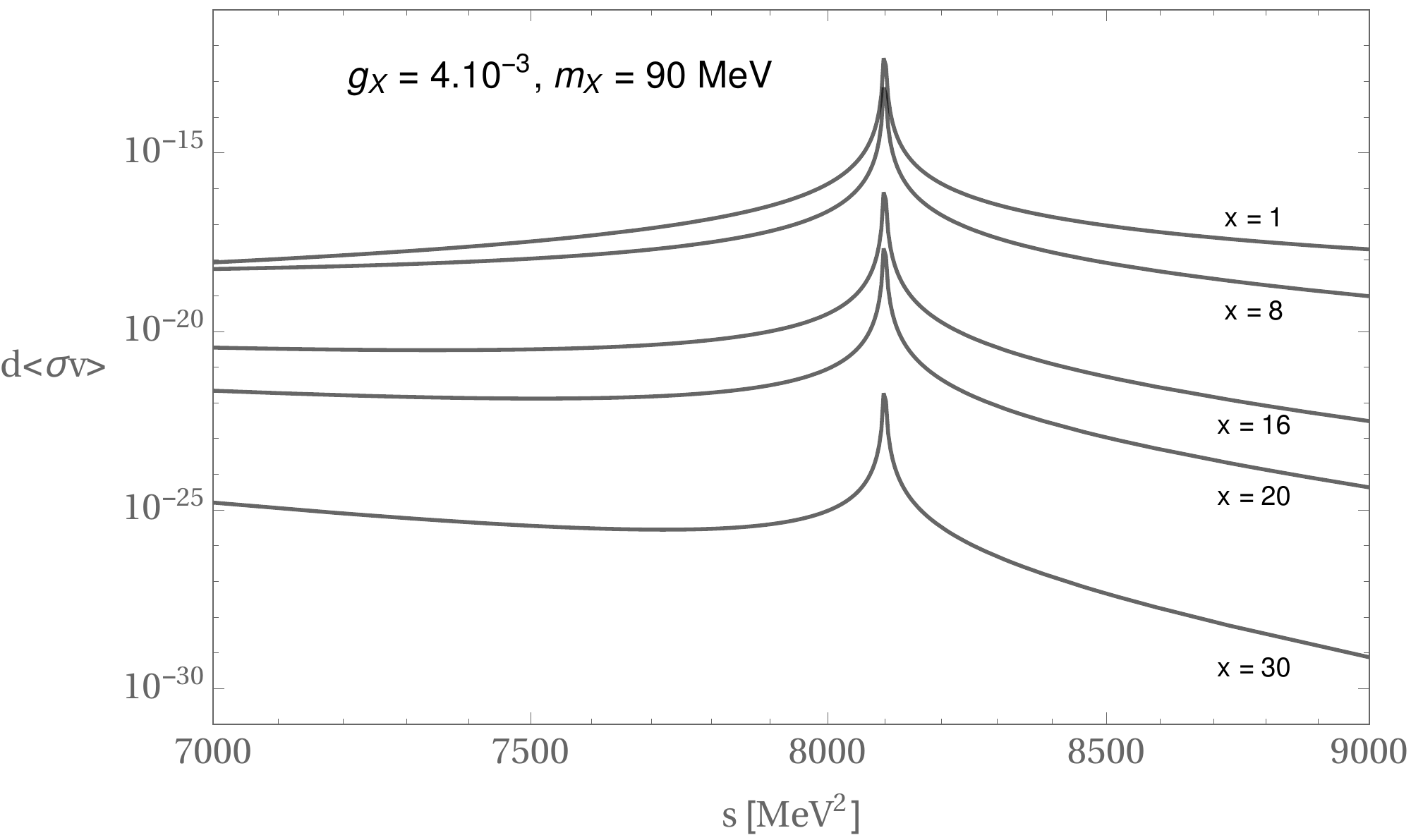}} 
		\\
		\subfloat[]{
		\includegraphics[width=10cm]{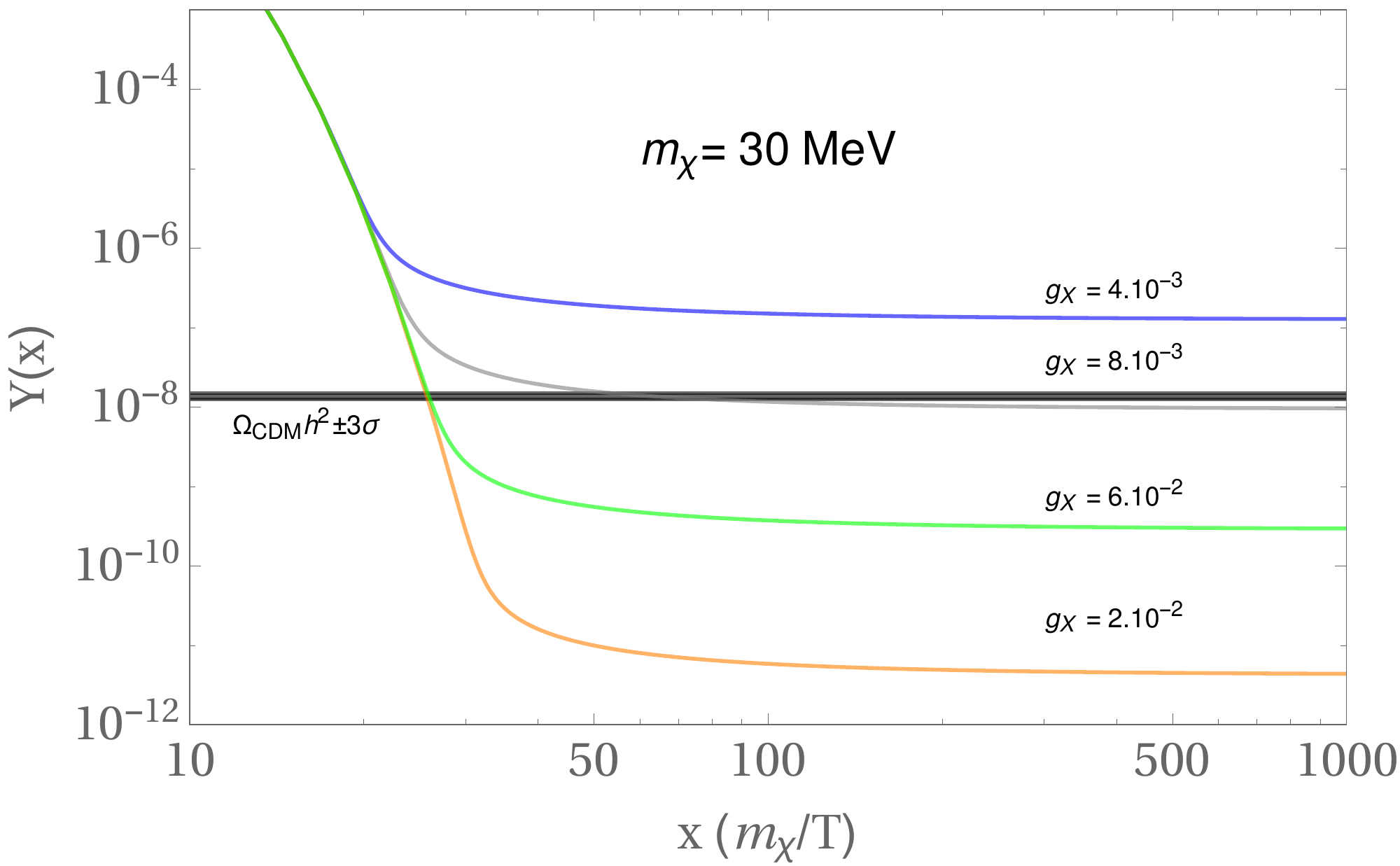}}
		\caption{(a) The differential thermal average dominated by a narrow resonance. In the example, the X boson mass is $m_X = 90$ MeV, while the dark fermion mass is $m_\chi = 30$ MeV. In (b), the horizontal black  band presents the $3\sigma$ region allowed by the current measurement of cold dark matter density. The abundance $Y_0$ is sensitive to small variations of the coupling constant $g_X$, such that $\Omega_{CDM}$ provides a strong bound for $\text{SM}\otimes U(1)_X$ theories. The remaining parameters are fixed as $(c_\beta, \kappa, F_{\mu \mu}) = (0.6, 1.5 c_\beta^2 g_X, 1)$}\label{fig:RA}
\end{figure}

\begin{figure}[tbp]
	\centering 
	\subfloat[]{ \includegraphics[width=.65\textwidth]{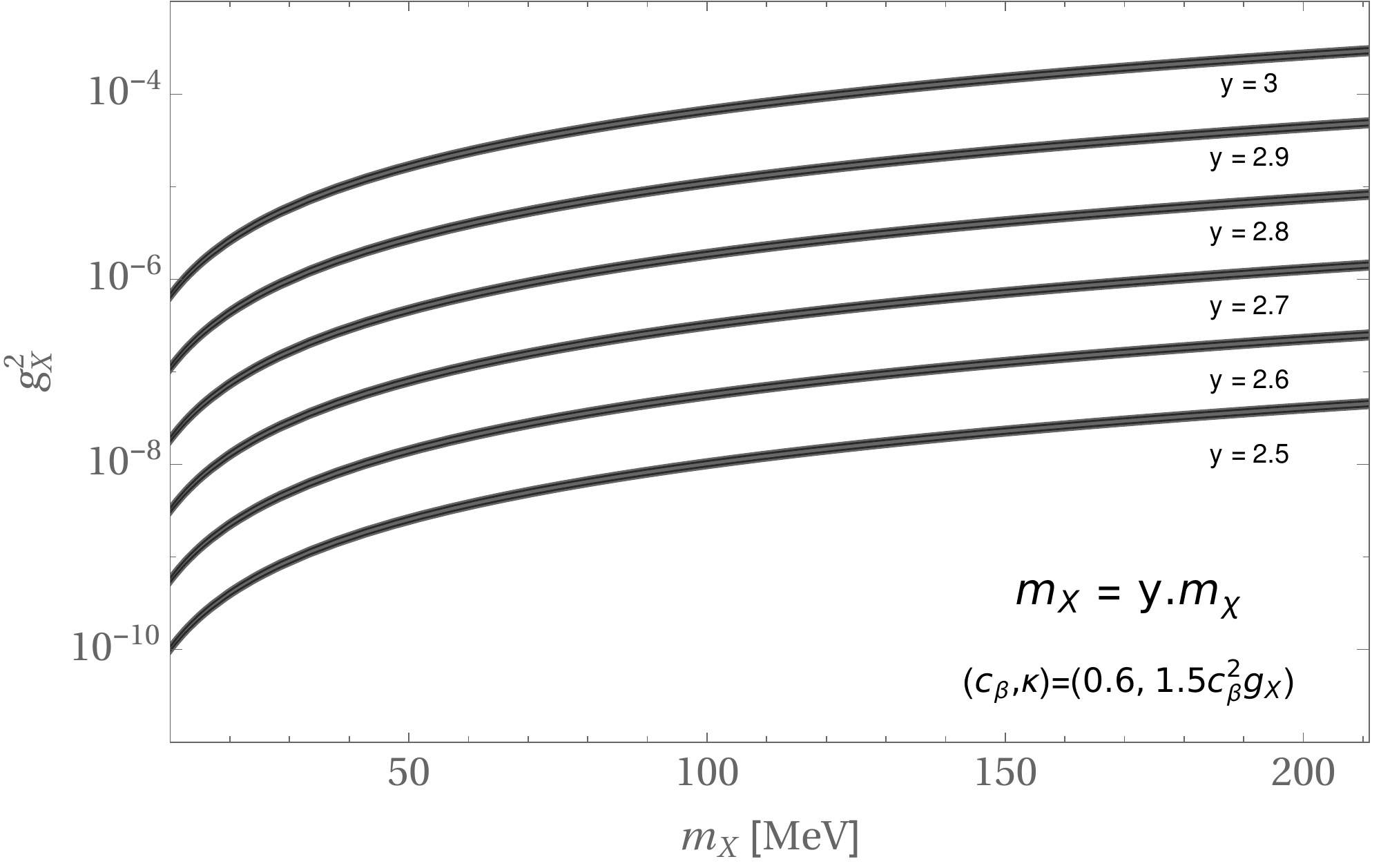}}
	\hspace{0.3cm}
	\subfloat[]{
	\includegraphics[width=.65\textwidth]{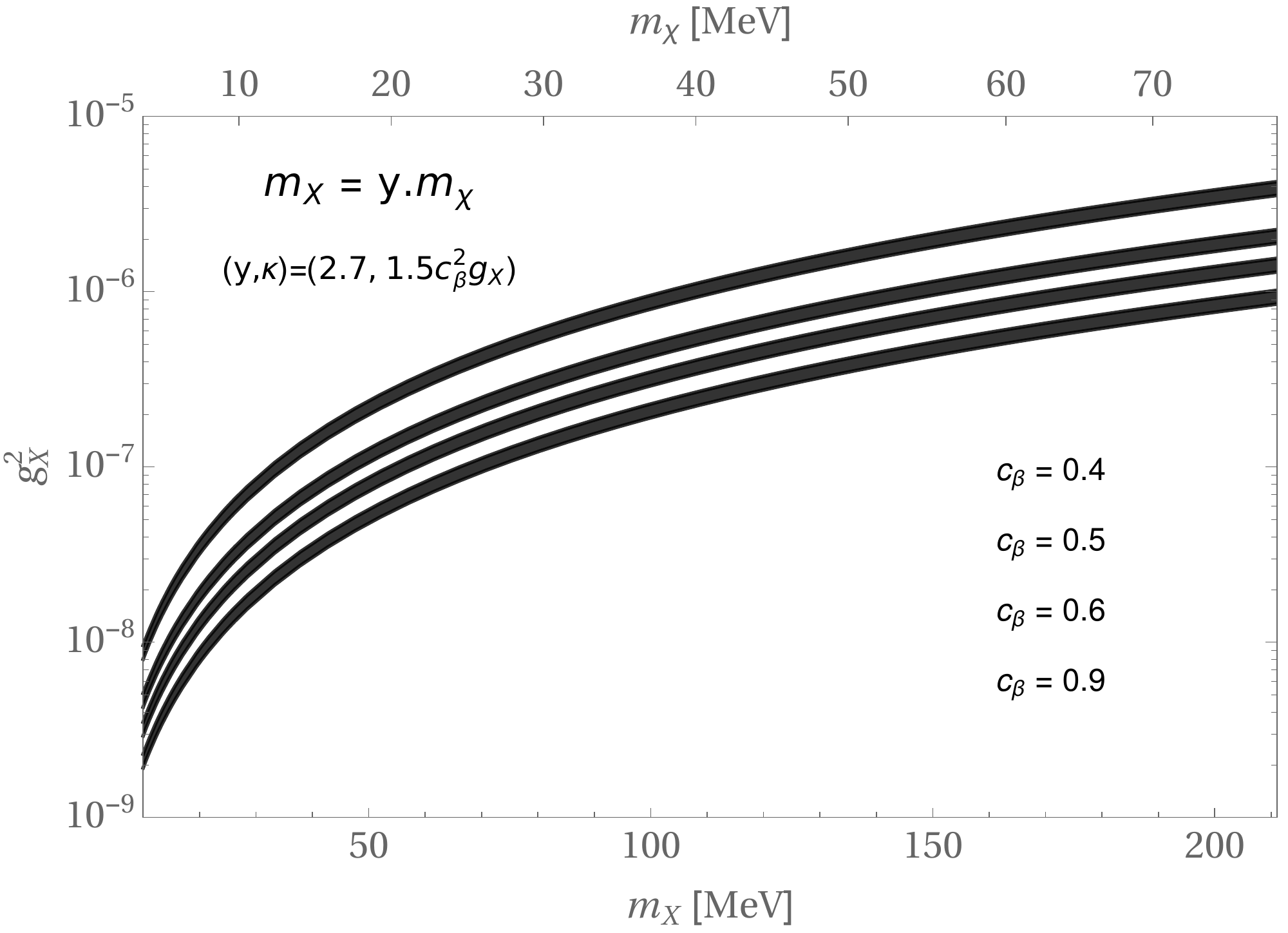}}
	\caption{\label{fig:RAchi} The $3\sigma$ favored region for $\Omega_{CDM}h^2$. In (a) it is illustrated how the variations in the relation $m_X = y m_\chi$ move the allowed region. In (b), the lines are less sensitive under small variations of $c_\beta$. The region below each line corresponds to overabundant $\Omega_\chi h^2$.}
\end{figure}

The thermally averaged annihilation cross-section is regulated by the diagram of Fig.\ref{fig:AFB} with $\chi \bar{\chi}$ in the initial states and $f$ summed over the particles at chemical equilibrium. Co-annihilations are discarded. The expression for  $\langle \sigma v \rangle$ can be written as \cite{Gondolo:1990dk}
\begin{equation}\label{Eq.therm}
\langle \sigma v \rangle = \frac{x}{8 m_\chi^5 k^2_2(x)} \int_{4 m^2_\chi}^\infty ds \left[\sqrt{s}(s-4m^2_\chi) k_1\left(x \frac{\sqrt{s}}{m_\chi}\right) \sigma \right], 
\end{equation}
where, as in Eq.(\ref{Eq.Yeq}), the $k_i$ are the order $i$ modified Bessel functions. The cross-section $\sigma$ is a sum over $f$ such that
\begin{equation}
d\sigma_f = \frac{d \Omega}{64 \pi^2} \frac{(s - 4m_f^2)^{1/2}}{(s - 4m_\chi^2)^{1/2}}\frac{\frac{1}{4}|\mathcal{M}_{\chi \bar{\chi}\rightarrow f \bar{f}}|^2}{s}
\end{equation}
is the differential cross section for each fermion $f$ in the final state. Again, the $X_\mu$ resonance is dominant over the remaining mediators and is implemented as a Breit-Wigner vector whose $\chi$ couplings are $x_V^\chi = x_A^\chi = g_X$. The total amplitude squared can be integrated over the polar angle and results in the expression 
\begin{equation}\footnotesize
\int d\theta |\mathcal{M}|^2 = g_X^4 \frac{8 m_\chi^2 \left[m_f^2(35 {x_A^f}^2 + 29 {x_V^f}^2) - 8s ({x_A^f}^2 + {x_V^f}^2)\right] + 19 s {x_A^f}^2 (s-4m_f^2) + s {x_V^f}^2 (19s - 52 m_f^2)}{3(\Gamma_X^2 m_X^2 + (m_X^2 -s)^2)}. 
\end{equation}
The solution of Eq.(\ref{Eq.Y}) then  has  to be translated in the present time. The abundance $Y_0$ is related to the WIMP relic density through \cite{Bertone:2010zza}
\begin{equation}
\Omega_\chi h^2 = 2.755 \times 10^5 Y_0 \frac{m_\chi}{MeV}
\end{equation} 
and must be consistent with the current measurement $\Omega_{CDM}h^2 = 0.1131(34)$ \cite{Hinshaw:2008kr} of the cold dark matter density. In Fig.\ref{fig:RA}(a) we present  the integrand of the thermal average in Eq.(\ref{Eq.therm}) for different temperatures, under a fixed point in the parameter space, dominated by the resonance\footnote{The numerical integration can be optimized by hiding the pole via the approximation
	\begin{equation}
	\int_{\alpha}^{\infty} f(x) dx \approx \int_{\alpha}^{x_p - \epsilon} f(x) dx + \int_{x_p + \epsilon}^{\infty} f(x) dx + \epsilon [f(x_p) + f(x_p - \epsilon)] ,
	\end{equation}
for a Breit-Wigner function whose pole is $x_p \gg \epsilon$.}. In Fig.\ref{fig:RA}(b) we  present the solution of Eq.(\ref{Eq.Y}) for the WIMP mass $m_\chi = 30$ MeV, $g_X= 4\times 10^{-3}$ and a set of couplings $x$. It depicts the point where the low temperatures hinder the abundance to follow the evolution of its equilibrium value, and the particle decouples.
The horizontal black  band is the $3\sigma$ limit for the current relic abundance. 
The well-known pattern of the Fig.\ref{fig:RA}(b) reveals that the relic density gets overabundant for small couplings, a feature that will help us to maximize the excluded region of our parameter space. Although an underabundant sector is disfavored by $\Omega_{CDM}$, it actually informs us about the necessity of completing the theory with additional dark matter candidates, and is not entirely ruled out. One can also notice from the figure how sensitive $Y_0$ is to small variations of $g_X$. We consider that the use of Eq.(\ref{Eq.RAap}) is appropriate for our precision level, and it may simplify our analysis. For large temperatures ($x < 20$) and small couplings ($g_X \lesssim 10^{-2}$, $c_\beta \in [0.4,0.95]$) the integral is dominated by the region around the resonance and is sufficiently narrow to use a Dirac delta approximation.\footnote{I.e. we replace the Breit-Wigner propagator by \begin{equation}
	\frac{1}{(s - m_X^2)^2 + (m_X \Gamma_X)^2} \rightarrow 
	\frac{\pi}{m_X \Gamma_X} \delta(s - m_X^2)
	\end{equation}} 
In order to illustrate the importance of the relation between  the $X_\mu$ and $\chi$ fermion masses, in  Fig. \ref{fig:RAchi}(a) we present the favored band for the current relic abundance by taking the relation $m_X = y m_\chi$ for different values of $y$, at the freeze out temperature $x_{f.o.} = 20$. In Fig. \ref{fig:RAchi}(b) a similar set of lines is computed for particular choices of $c_\beta$.  

\begin{figure}[tbp]
	\centering %
	\subfloat[]{
		\includegraphics[width=.65\textwidth]{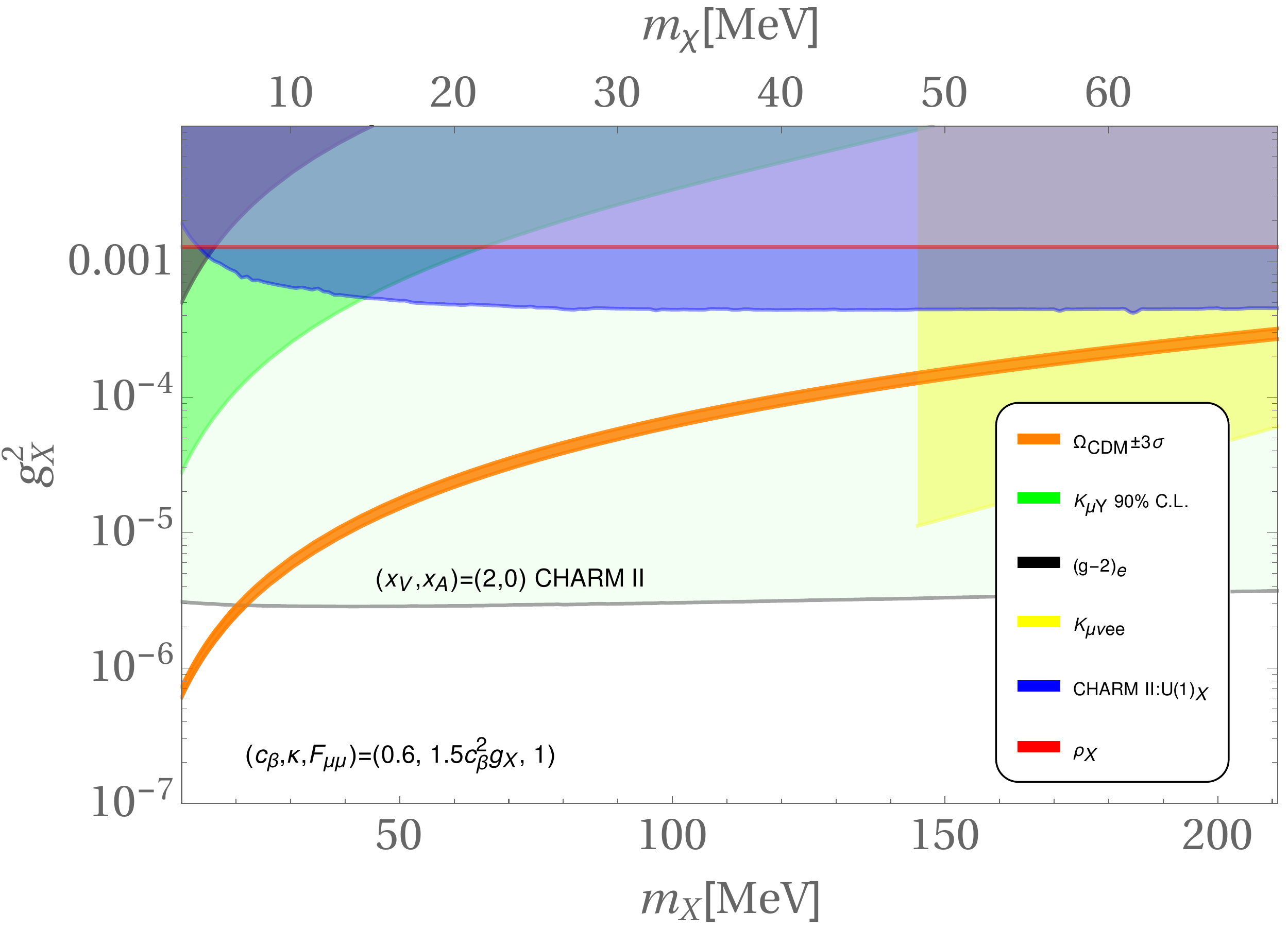}}
	\\
	\subfloat[]{
		\includegraphics[width=.65\textwidth]{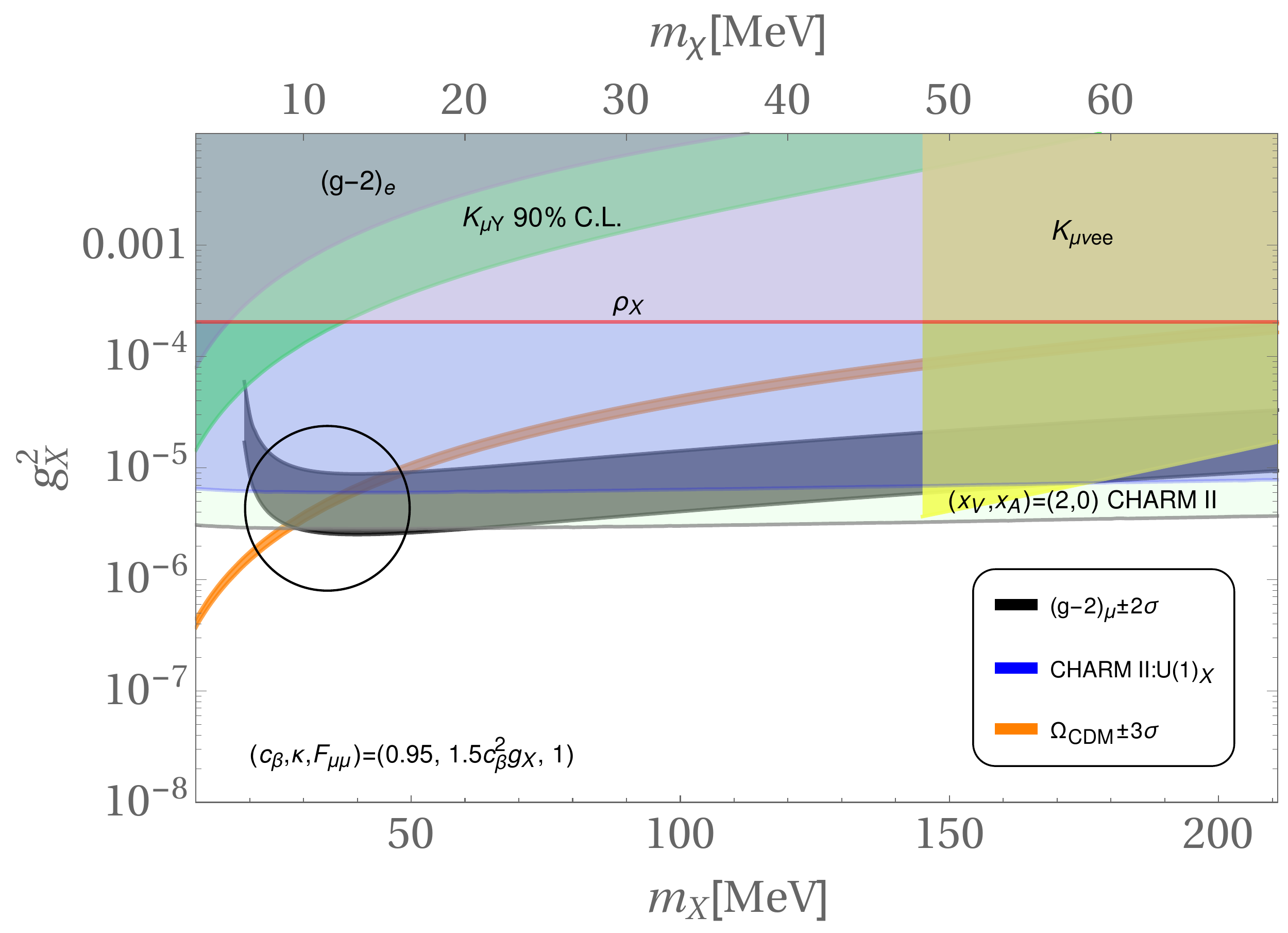}}
	\caption{\label{fig:Const} The parameter space for $\kappa = \frac{3}{2} c_\beta^2 g_X$. Notice, in (a), the excluded region for the dark photon ($A'$ )(light-blue) is presented in comparison to the dark gauge boson  ($Z'$)  region  (dark-blue). In (b), the difference is reduced when $c_\beta = 0.95$ which, on the other hand, produces a possible solution both for the $(g-2)_\mu$ discrepancy and the $\Omega_{CDM}$ (circled region).} 
\end{figure}

\section{Outlook}
The results of the previous section, for two fixed points, were summarized in Fig.\ref{fig:Const}. In plot (a), $c_\beta = 0.6$ cannot resolve the $(g-2)_\mu$ discrepancy. Notwithstanding, we can verify the impact of the axial-vector couplings interference in the cross-section of the neutrino trident production   (dark-blue region) in comparison with the dark photon case (light-blue). Moreover, the choice of the parameter  $\kappa = \frac{3}{2} c_\beta^2 g_X$ renders irrelevant the bound coming from 
$(g-2)_e$. In  plot (b) the favored region for the explanation of the $(g-2)_\mu$ discrepancy (in gray) is presented, for $c_\beta = 0.95$, while the circled area highlight the overlap between the solution for the muon anomaly and the relic abundance $\Omega_{CDM}$ favored bands. 
 The parameter space that we covered hitherto should be further tested. In the following subsection  \ref{Sec:Out}, we  consider the forward-backward asymmetry in  electron-positron collision to fermion anti-fermion, $e^+ e^- \rightarrow \bar{f} f$, in order to determine how the observable is sensitive on the $\text{SM}\otimes U(1)_X$ parameters. Finally, in subsection \ref{Sec:Mmu2ee} we illustrate the impact of one set of parameters, allowed by Fig. \ref{fig:Const}(b), to the leptonic decays $M \rightarrow j \nu_j i^+ i^-$, where $M = \pi, K, D, D_s, B$ and $i,j = e, \mu$.

\subsection{Parity Non-Conserving Observables}\label{Sec:Out}

\begin{figure}[tbp]
	\centering 
	\includegraphics[width=7cm]{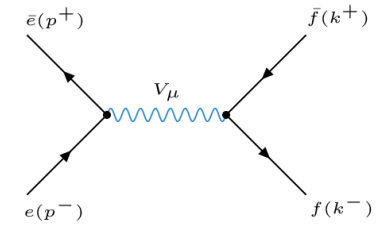}
	\caption{\label{fig:AFB} Forward-backward asymmetries in $e^+ e^- \rightarrow \bar{f} f$ are important tests for axial-vector couplings. The model predict non-universality for $f = \mu$ and $f = \tau$. Here $V = \gamma, Z, X$.}
\end{figure}

Light $Z'$ physics requires the use of  the LEP data,  where the effects of $Z$ interactions are suppressed by its large mass.  Here, we have to constrain the region for the 
light $Z'$ and its small couplings. As  we mentioned in the previous section, in  scattering processes, the  dark photon effects can be considered as a correction to the fine-structure constant.  When the axial-vector coupling of the dark gauge boson is  present it can be tested  in the angular asymmetries of the differential cross-section. The forward-backward asymmetry is defined as 
\begin{equation}
A(\theta) \equiv \frac{d\sigma(\theta) - d\sigma(\pi - \theta)}{d\sigma(\theta) + d\sigma(\pi - \theta)}. 
\end{equation}
Here we will focus on the energy region far from both $Z$ and $X$ peaks, i.e. $2m_\mu \ll \sqrt{s} \ll m_Z$ and we must compute the generic diagram in  Fig.\ref{fig:AFB} for $V = \gamma, Z, X$. For simplicity, the Feynman rule for the vertex $\bar{f} f V_\mu$ can be  written as $i e \gamma_\mu (v_f^V - a_f^V \gamma_5)$. For instance, $(v_f^\gamma, a_f^\gamma) = (-q_f, 0)$ where $q_e = -1, q_u = \frac{2}{3}, q_d = - \frac{1}{3}$. The amplitude can  be expressed as 
\begin{equation}
\mathcal{M}_V = \frac{e^2}{s - m^2_V} 
[\bar{v}(p^+) \gamma^\mu (v_e^V - a_e^V \gamma_5) u(p^-)][\bar{u}(k^-) \gamma_\mu (v_f^V - a_f^V \gamma_5) v(k^+)], 
\end{equation}
with $|\mathcal{M}|^2 = |\sum_{V = \gamma, Z, X} \mathcal{M}_V|^2$. The dominant contribution to $A(\theta)$ originates from the interference of contributions coming from $\gamma Z$ and $\gamma X$ in the numerator, while the cross section coming from the photon mediator ($\gamma$) gives the dominant  contribution in the denominator, i.e
\begin{equation}
A(\theta) \approx \frac{[d\sigma^{\gamma Z}(\theta) + d\sigma^{\gamma X}(\theta)] - [d\sigma^{\gamma Z}(\pi - \theta) + d\sigma^{\gamma X}(\pi - \theta)]}{d\sigma^\gamma (\theta) + d\sigma^\gamma (\pi - \theta)}.
\end{equation}
In the CM reference frame it results in
\begin{equation}
A(\theta) \approx \frac{8 s c_\theta |\mathbf{k}| \sqrt{s}}{4 c_\theta^2 |\mathbf{k}|^2 + 4m_f^2 + s} \left[\frac{a_e^X a_f^X}{s - m^2_X} + \frac{a_e^Z a_f^Z}{s - m^2_Z}\right]. 
\end{equation}
Here $c_\theta$ is the scattering angle, and $\mathbf{k}$ the space-momentum of the particles in the final state.  In the region $\sqrt{s} \gg m_\mu$ the contribution coming from $X$ exchange can be written as  $\delta A^X(\theta) \propto \frac{a_e^X a_f^X}{s}$ or
\begin{equation}
A(\theta) \propto \left[\frac{a_e^X a_f^X}{s} - \frac{a_e^Z a_f^Z}{m^2_Z}\right]. 
\end{equation} 
In our case, by assuming $a^Z \sim g$ and $a^X \sim g_X$, the region around $\sqrt{s} \sim \frac{m_Z}{10}$ would severely constrain $g_X \sim 10^{-1} g$. 
In summary, the precise measurement of forward-backward asymmetry, in the region described above, might provide an additional limit to the coupling $g_X$, as well as a possible test of lepton-flavor universality. 

\subsection{Leptonic Meson Decays: \texorpdfstring{$M\to l' \nu_{l'} ll$}{}}\label{Sec:Mmu2ee}

\begin{figure}[tbp]
	\centering 
	\subfloat[]{ \includegraphics[width=.75\textwidth]{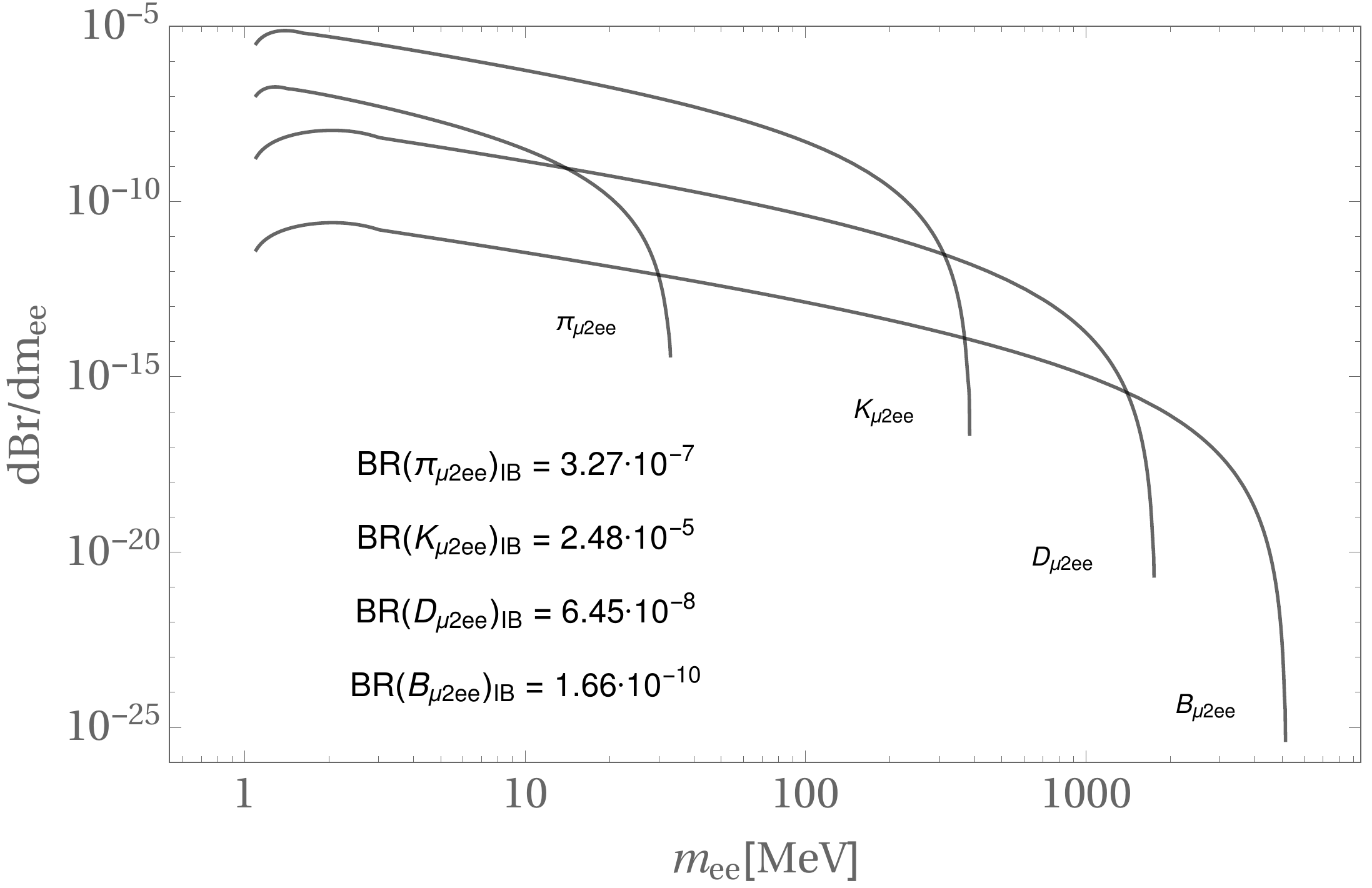}}
	\hspace{0.3cm}
	\subfloat[]{
	\includegraphics[width=.75\textwidth]{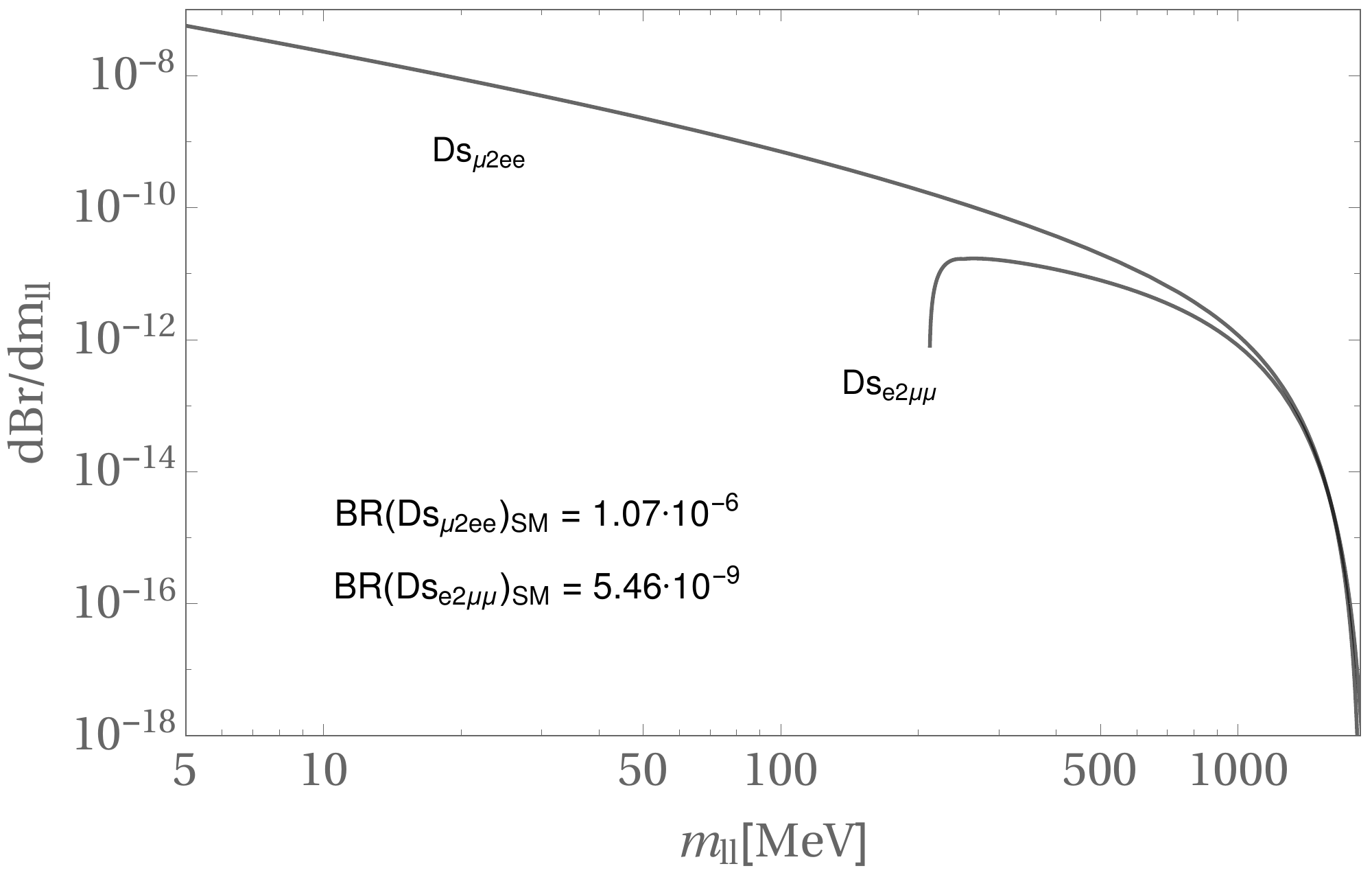}} 
    \caption{\label{fig:IB} Differential branching ratio as a function of  the di-lepton invariant mass in the SM. In  the plot (a), the IB diagrams are dominant. In (b), the IB and SD contributions are presented for $D_s$ decays.}
\end{figure}

The experimental measurement of the branching ratio for $K_{\mu \nu ee}$ gives the most stringent bound of Fig.(\ref{fig:Const}). This result motivates us  to extrapolate this analysis and  make predictions for the bottom and charm mesons. In this section we present  our results for the total and differential branching ratio of the 
purely leptonic decays $M \rightarrow l' \nu_{l'} l^+l^- $, or simply $M_{l'2ll}$, where $M$ denotes the mesons $M = \pi, K, D, D_s, B$ and $l',l = e, \mu$. The diagrams in  Fig.\ref{fig:KmuX} are dominant in that  process and can be separated into the inner-bremsstrahlung (IB) amplitude (a,b) and a structure dependent (SD) amplitude (c). The off-shell vectors $X_\mu, \,A_\mu$ mediate the interaction with the lepton pair $l^+ l^-$. The IB amplitudes can be computed in a general form for all mesons by following the work of \cite{Bijnens:1992en}. The SD diagram can be parametrized as
\begin{equation}
\mathcal{M}_{SD} = e^2 \frac{G_F}{\sqrt{2}} V_{UD}^* h_{\mu \alpha} [\overline{u}_\nu \gamma^\mu (1-\gamma_5)v_{l'}][\overline{u}_il\gamma^\alpha v_l] ,
\end{equation} 
where $V_{UD}$ is the CKM element linked to the particular meson's quark structure, and 
\begin{equation}
h_{\mu \alpha} = \frac{1}{p^2} \left[\epsilon_{\mu \alpha \lambda \beta} k^\lambda p^\beta \frac{F_V(p^2)}{m_M} - ie \left(g_{\mu \alpha} p \cdot (k - p) - p_\mu(k - p)_\alpha \right)\frac{F_A(p^2)}{m_M}\right].
\end{equation}
In general, the IB contribution is dominant in $M_{\mu 2 ee}$ and suppressed in $M_{e 2 \mu \mu}$. We compute the  differential branching ratio for $D_s$ decay, in the SM framework, including both IB + SD interference. The form-factors are given by\footnote{We acknowledge D. Melikhov for providing us with these form-factors.}:
\begin{equation}
F_V(p^2) = \frac{0.029}{1 - \frac{p^2}{(2.11 GeV)^2}} ,   \qquad 
F_A(p^2) = \frac{0.173}{1 - \frac{p^2}{(2.46 GeV)^2}}.
\end{equation}
In Fig.\ref{fig:IB} we plot the results for different mesons, keeping IB contributions only. 
The results are summarized in Table  \ref{tab:i}. We provide cuts on the di-lepton invariant mass in order to connect them to possible experimental limitations. These results for the SM branching ratio might be interesting if precision measurements are reached. 
In  Fig.\ref{fig:IB}(b) we present  the results for the $D_s$ meson including the SD  part. In these last examples, the di-leptons are the $e^+e^-$ and $\mu^+\mu^-$ pairs and the integrated values are presented in Table \ref{tab:ii}. 

\begin{figure}[tbp]
	\centering %
	\includegraphics[width=.75\textwidth]{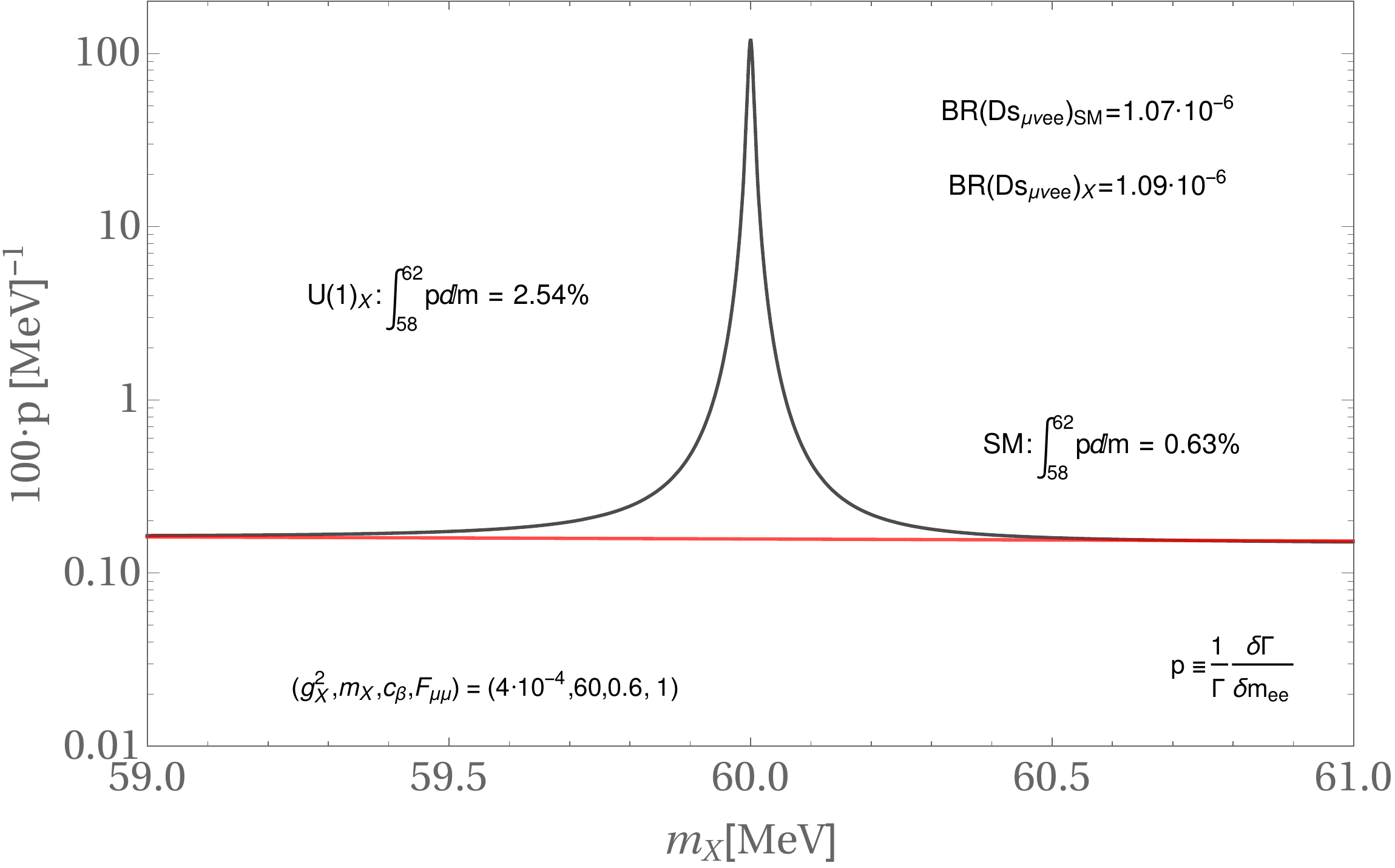}
\caption{\label{fig:Ds} The normalized differential branching ratio corresponds to the probability $P = 2.54\%$ of measuring the di-lepton mass in the interval $58$ MeV $ < m_{ee} < 62$ MeV at the resonance $(g_X ^2, m_X) = (4 \times 10^{-4}, 60)$, one allowed region from Fig.\ref{fig:Const} (a). $P = 0.63\%$, in the SM framework.} 
\end{figure}

We also compute the branching ratios for $M_{\mu 2 ee}$  ($M=\pi,\, K, \, D,\, B$), coming from the  dominant IB amplitudes, in which the photon mixes with the X boson.  The branching ratios below test our IB formulas in the $\text{SM}\otimes U(1)_X$ framework for the point of Fig.\ref{fig:Ds}, namely $(g_X ^2, m_X) = (4 \times 10^{-4}, 60)$: 
\begin{equation}
\begin{matrix}
\text{Br}(\pi_{\mu2ee})_{IBX} = 3.27 \times10^{-7},& & \text{Br}(K_{\mu2ee})_{IBX} = 2.49 \times10^{-5}, \\
\text{Br}(D_{\mu2ee})_{IBX} = 6.54 \times 10^{-8}, & & \text{Br}(B_{\mu2ee})_{IBX} = 1.78 \times 10^{-10},\\
\end{matrix}
\end{equation}
Once more, the increase in the differential branching ratios is tinny and it can eventually be observed if high precision measurements are performed.

Finally, in order to provide a possible test for lepton flavor universality we introduce 
	\begin{equation}
	R(f) = \frac{Br(M_{f2\mu \mu})}{Br(M_{f2e e})} ,
	\end{equation} 
 in a kinematic region far from resonances. In the SM the ratio is close to the unity for $q^2 >>  (2m_\mu)^2$ and different leptons in the final state. This case might be potentially interesting, and it is kinetically allowed only for $B_{\tau 2 ll} $.
	For $m_{ll}^2$ far from the $X_\mu$ pole, any non-universality effects are negligible if compared with the SM prediction. For instance, from the parameters selected for the analysis presented in  Fig.\ref{fig:Ds}, and $(1500)^2$ MeV$^2$  $< m_{ll}^2  < (1600)^2$ MeV$^2$, we find
	\begin{align}
	R_X(\tau) = \begin{matrix}
	0.93 \qquad &m_X = 1550 \ \text{MeV}, \\ 
	0.99  \qquad &m_X = 60 \ \text{MeV}, 
	\end{matrix}
	\end{align}    
while for the SM value,  $R_{SM}(\tau) = 0.9998$. In the region where the invariant mass  of the  di-lepton pair is in $(300)^2$ MeV$^2$  $< m_{ll}^2  < (400)^2$ MeV$^2$ 
we find 	
	\begin{equation}
	R_{SM}(\tau) = 0.933, \qquad R_X(\tau) = \begin{matrix}
	0.90 \qquad m_X = 350 \ \text{MeV},\\ 
	0.931  \qquad m_X = 60 \ \text{MeV}. 
	\end{matrix}
	\end{equation}   
	
	Finally, in  Fig.\ref{fig:Ds} we present the normalized differential branching ratio for $Ds_{\mu2ee}$, both in the $\text{SM}\otimes U(1)_X$ and in the SM frameworks. Around the resonance, the probability for measuring the di-lepton mass in the interval $58$ MeV $ < m_{ee} < 62$  MeV is equal to $P = 2.54\%$, for $(g_X ^2, m_X) = (4 \cdot 10^{-4}, 60)$, allowed by Fig.\ref{fig:Const}(a), in comparison with $P = 0.63\%$, in the SM.

\begin{table}[tbp]
	\centering
	\begin{tabular}{|c|llll|}
		\hline
		$M_{\mu2ee}$ & Full & $m_{ee} \geqslant 0.1m_M$ & $m_{ee}\geqslant 20$MeV & $m_{ee}\geqslant 140$MeV \\
		\hline 
		$\pi_{\mu2ee}$ & $3.27 \times 10^{-7}$ & $2.82 \times 10^{-9}$ & $-$ & $-$
		\\
		 $K_{\mu2ee}$ & $2.48 \times 10^{-5}$ & $8.6 \times 10^{-7}$ & $3.15 \times 10^{-6}$ & $4.97 \times 10^{-8}$
		\\
		$D_{\mu2ee}$ & $6.45 \times 10^{-8}$ & $1.13 \times 10^{-9}$ & $1.45 \times 10^{-8}$ & $1.84 \times 10^{-9}$
		\\
	    $B_{\mu2ee}$ & $1.66 \times 10^{-10}$ & $1.64 \times 10^{-12}$ & $4.69 \times 10^{-11}$ & $1.02 \times 10^{-11}$
		\\
		\hline
	\end{tabular}
	\caption{\label{tab:i} In the SM framework, a set of cuts for the di-lepton mass for the $BR(M_{\mu2ee})$, used in the decay width calculations for IB contributions. }
\end{table}

\begin{table}[tbp]
	\centering
	\begin{tabular}{|c|llll|}
		\hline
		$M_{f2ll}$ & Full & $m_{ll} \geqslant 0.1m_{Ds}$ & $m_{ll}\geqslant 20$MeV & $m_{ll}\geqslant 140$MeV \\
		\hline 
		$Ds_{\mu2ee}$ & $1.07 \times 10^{-6}$ & $2.41 \times 10^{-8}$ & $2.56 \times 10^{-7}$ & $3.95 \times 10^{-8}$
		\\
		$Ds_{e2\mu\mu}$ & $5.46 \times 10^{-9}$ & $-$ & $-$ & $-$
	    \\
		\hline
	\end{tabular}
	\caption{\label{tab:ii} In the SM framework, similar cuts for the di-lepton mass for the $BR(Ds_{j2ii})$, used in the decay width calculations for IB+SD  contributions.}
\end{table}

\section{Conclusion}\label{Sec.Coc}
We have explored  a number of processes  at low energies (MeV-GeV) 
in order to constrain a $\text{SM}\otimes U(1)_X$ theory, UV-completed by cold WIMPs and  whose $U(1)_X$ is chiral for  right-handed fermions. The most stringent bounds are obtained from the relic abundance accompanied  by the electron anomalous magnetic moment and the neutrino trident production. 
The dark photons, as  $U(1)_X$  gauge bosons, may be fully excluded from these observables.
On the other hand, the theories where the  $X_\mu$ bosons are coupled to the fermion fields through both vector and axial-vector currents ($Z'$ bosons)
are not in conflict with the li\-mits from the neutrino trident production, but can still be ruled out by $(g-2)_e$. However, a possible interference between the vector and axial-vector couplings might make the shift in the fine structure constant negligible. Under this condition a certain choice of parameters can be found for three tuned planes, which allow the explanation of the $(g-2)_\mu$ discrepancy for specific values of the $c_\beta$ angle. In particular, the pairs $\kappa = -\frac{7}{5}$ for $c_\beta > 0.7$, $\kappa = 3$ for $c_\beta > 0.99$ and $\kappa = \frac{3}{2}$ for $c_\beta > 0.95$, solve the discrepancy in the muon system without the introduction of light scalars.
 
We propose to measure the forward-backward asymmetry in $e^+ e^- \rightarrow \bar{f} f$, for $f = \mu, \tau$, far from the  $Z$ boson peak at low energies. We also suggest the measurement of the branching ratios to the purely leptonic meson decays $M\to j \nu ll$, for $M = D, D_s, B$. The ratio $BR(M\to j \nu_j \mu^+ \mu^-) /BR(M\to j \nu_j e^+ e^-)$ in definite regions of the di-lepton invariant mass 
shows a deviation from the SM prediction and might serve as an important test of lepton flavour universality, in particular for $M = B$ and $j = \tau$. 
\newpage
\appendix

\acknowledgments
We are grateful to Dmitri  Melhikhov for providing us with the form-factors present in the $D_s\to \mu\nu ee$ amplitude.  
S.F. acknowledges support of the Slovenian Research Agency under the core funding grant P1-0035. F.C.C. would like to thank Clara H. Feliu, prof. Gudrun Hiller, Peter Schuh, Dennis Loose, Mathias Becker, Andrey Tayduganov, prof. Emmanuel A. Paschos and prof. Heinrich P\"as for useful discussions. F.C.C acknowledges support from the BMBF grant ``Verbundprojekt 05H2015:  Quark-Flavor-Physik am LHC (BMBF-FSP 105), Flavorsignaturen in Theorie und Experiment - LHCb: Run 2 and Upgrade'' and from the Technische Universit\"at Dortmund, Department of Physics.
 
\section{Appendix}
\subsection{Feynman rules}\label{Ap.FR}
The Feynman rules for the vertexes presented Fig.\ref{fig:KmuX}(a,b) may be  generically written as
\begin{eqnarray}
M^+(k) \ \bar{\nu}_l(p_i) \ l(p_j) &:& -i \frac{G_F}{\sqrt{2}} f_M V^*_{UD} \ \slashed k (1-\gamma_5), \\
X_ \mu(p_m) \ \bar{l} (p_i) \ l(p_j) &:& i \frac{\gamma_\mu}{2} (x_V^l + x_A^l \gamma_5) \label{Eq.evert},\\
X^\mu (p_m) \ M^+(p_i) \ M^-(p_j) &:& - i c_\phi^2 \kappa \ (p_i^\mu + p_j^\mu).
\end{eqnarray}

\subsection{Decay Width}\label{AP.DW}
The general expression for the decay width can be written as 
\begin{equation}\label{Eq.DecayFormula}
d \Gamma = \sum_{s p} \frac{|\mathcal{M}|^2}{2 M} \frac{\Phi_n (M; m_1, \cdots, m_n)}{(2 \pi)^{(3n - 4)}}, 
\end{equation}
such that, for $n=3$,  following  results of Ref. \cite{Kersevan:2004yh}, 
\begin{equation}\label{Eq.Phi3}
\Phi_3 (M; m_1, m_2, m_3) =  dM_2^2\Bigg|_{(m_1 + m_2)^2}^{(M - m_3)^2} \frac{\sqrt{\lambda(M^2, M_2^2, m_3^2)}}{8 M^2} d\Omega^*_3 \frac{\sqrt{\lambda(M_2^2, m_1^2, m_2^2)}}{8 M_2^2} d\Omega^*_2 . 
\end{equation} 
The variables defining the solid angle $d\Omega^*_i$ are in the rest frame of $k_i = \sum_{j=1}^{i} p_j$,   $M_i^2 \equiv k_i^2$.  
The scalar products emerging from the squared amplitudes can be expressed in a more convenient form through the momenta $k_i$, i.e.
\begin{equation}
k_1^2 = 0, \qquad k_2^2 = M_2^2, \qquad k_3^2 = M^2
\end{equation}
and 
\begin{subequations}
	\begin{eqnarray}
k_1 \cdot k_2 &=& \frac{M_2^2 - m^2_X}{2} , \enspace  k_2 \cdot k_3 =\frac{M^2 + M_2^2 - m^2_l}{2} \\
k_1 \cdot k_3 &=& \frac{(k_1 \cdot k_2)(k_2 \cdot k_3)}{M_2^2} - \frac{(k_1 \cdot k_2) \sqrt{(k_2 \cdot k_3)^2 - M^2 M_2^2}}{M_2^2} c_\theta^*. 
\end{eqnarray}
\end{subequations}
Using these invariants it follows that $q^2 =m^2_X + m^2_l - M_2^2 + 2 k_1 \cdot k_3$ and $q_{23}^2 = M^2 - 2k_1 \cdot k_3$. 
\subsection{Phase space integration for the neutrino trident production}
The integration  limits for $t_i$ are obtained  from the condition
\begin{equation}
\begin{aligned}
t_2 &: G(s_2, t_2, m_3^2, t_1, m_b^2, m_2^2) \leq 0 \\
t_1 &: G(s, t_1, s_2, m_a^2, m_b^2, m_1^2) \leq 0
\end{aligned}
\end{equation}
where $G$ is the Cayley determinant:
\begin{equation}
G(x,y,z,u,v,w) = -\frac{1}{2} 
\begin{vmatrix}
0 & 1 & 1 & 1 & 1 \\
1 & 0 & v & x & z \\
1 & v & 0 & u & y \\
1 & x & u & 0 & w \\
1 & z & y & w & 0
\end{vmatrix}. 
\end{equation}
All the possible scalar products in the real photon-neutrino scattering can be written in terms of the invariants of Eq.(\ref{EqNeuInv}) and it is  convenient to write $s_1$ in terms of the polar angle via
\begin{equation}
\begin{aligned}
s_1 &= s + m_3^2 - \frac{1}{\lambda(s_2,t_1,m^2_b)} 
\Biggl[ 
\begin{vmatrix}
2m_b^2 & s_2 - t_1 + m_b^2 & m_b^2 + m_3^2 -t_2 \\
s_2 - t_1 + m_b^2 & 2s_2 & s_2 - m_2^2 + m_3^2 \\
s - m_a^2 + m_b^2 & s + s_2 - m_1^2 & 0
\end{vmatrix} + \\
& + 2\left(G(s,t_1,s_2,m_a^2,m_b^2,m_1^2) G(s,t_1,s_2,m_a^2,m_b^2,m_1^2)\right)^{1/2} \cos \phi\Biggr].
\end{aligned}\end{equation}

The remaining angles in Eq.(\ref{Eq.Phi3}) may be integrated directly and $\Phi_3$ is reduced to
\begin{equation}\label{Eq.Phi3b}
\Phi_3 (M; m_1, m_2, m_3) = \frac{\pi^2}{4M^2} dM_2\Bigg|_{(m_1 + m_2)}^{(M - m_3)}
dc_\theta^* \Bigg|_{-1}^{1}  \frac{\sqrt{\lambda(M^2, M_2^2, m_3^2)} \sqrt{\lambda(M_2^2, m_1^2, m_2^2)}}{M_2} 
\end{equation} 







\providecommand{\href}[2]{#2}\begingroup\raggedright\endgroup

\end{document}